\documentclass[%
notitlepage,amsmath,amssymb,
]{revtex4-1}

\usepackage{graphicx}
\usepackage{dcolumn}
\usepackage{bm}

\graphicspath{{FIGURES/}}

\usepackage{color}

\newcommand{\bA}{{\boldsymbol A}}
\newcommand{\bB}{{\boldsymbol B}}

\newcommand{\bI}{{\boldsymbol I}}

\newcommand{\bQ}{{\boldsymbol Q}}
\newcommand{\bR}{{\boldsymbol R}}

\newcommand{\bT}{{\boldsymbol T}}

\newcommand{\ba}{{\boldsymbol a}}
\newcommand{\be}{{\boldsymbol e}}
\newcommand{\bg}{{\boldsymbol g}}
\newcommand{\bk}{{\boldsymbol k}}
\newcommand{\bfm}{{\boldsymbol m}}

\newcommand{\bq}{{\boldsymbol q}}
\newcommand{\bp}{{\boldsymbol p}}

\newcommand{\fbf}{{\boldsymbol f}}

\newcommand{\bv}{{\boldsymbol v}}

\newcommand{\bx}{{\boldsymbol x}}

\newcommand{\bX}{{\boldsymbol X}}
\newcommand{\bchi}{{\boldsymbol \chi}}
\newcommand{\bnu}{{\boldsymbol \nu}}
\newcommand{\bmu}{{\boldsymbol \mu}}

\newcommand{\bim}{{\boldsymbol \imath}}
\newcommand{\bjm}{{\boldsymbol \jmath}}
\newcommand{\bvep}{\delta\bv} 
\newcommand{\Cbeta}{{\mathcal C}_\beta}
\newcommand{\Vbeta}{{\mathcal V}_\beta}
\newcommand{\xbeta}{\bx_\beta}
\newcommand{\Ibeta}{I_\beta}
\newcommand{\nbeta}{n_\beta}

\renewcommand{\epsilon}{\varepsilon}
\renewcommand{\phi}{\varphi}

\newcommand{\transpose}[1]{{#1}^{\scriptscriptstyle\top\mskip-3.75mu}}

\begin{document}


\title[Spatial averaging of a dissipative particle dynamics model for active suspensions]{Spatial averaging of a dissipative particle dynamics model for active suspensions}

\author{Alexander Panchenko}
 \email{panchenko@math.wsu.edu.}
 \affiliation{Department of Mathematics, Washington State University, 
 Pullman, WA 99164, USA.}
\author{Denis F.\ Hinz}%
 \email{dfhinz@gmail.com.}
\affiliation{ 
Flow Laboratory, Kamstrup A/S, Industrivej 28\\
Stilling, 8660 Skanderborg, Denmark
}%

\author{Eliot Fried}
 \email{eliot.fried@oist.jp.}
\affiliation{%
Mathematics, Mechanics, and Materials Unit\\ Okinawa Institute of Science and Technology Graduate University\\ Onna, Okinawa 904-0495, Japan
}%

\date{\today}

\begin{abstract}
Starting from a fine-scale dissipative particle dynamics (DPD) model of self-motile point particles, we derive meso-scale continuum equations by applying a spatial averaging version of the Irving--Kirkwood--Noll procedure. Since the method does not rely on kinetic theory, the derivation is valid for highly concentrated particle systems. Spatial averaging yields a stochastic continuum  equations similar to those of Toner and Tu. However,  our theory also involves a constitutive equation for the average fluctuation force. According to this equation, both the strength and the probability distribution vary with time and position through the effective mass density. The statistics of the fluctuation force also depend on the fine scale dissipative force equation,  the physical temperature, and two additional parameters which characterize fluctuation strengths. Although the self-propulsion force entering our DPD model contains no explicit mechanism for aligning the velocities of neighboring particles, our averaged coarse-scale equations include the commonly encountered cubically nonlinear (internal) body force density.

%
\end{abstract}

\pacs{Valid PACS appear here}

\maketitle


\section{Introduction}

The study of active suspensions is by now a well established field of research in mathematics, physics, and the engineering sciences. Because the literature on active suspensions is vast, an exhaustive review of its contents is infeasible. We therefore include only a few representative citations of prior work \cite{bricard, ibele, koch-sub, kudrolli08, kudrolli10, narayan06, narayan07, schaller}.

In particular, concentrated suspensions of active particles have recently attracted much attention \cite{marchetti-review}.  For such media, the derivation of continuum theories from realistic fine scale models constitutes a challenging and still largely open problem.  Most available results rely on kinetic theory \cite{baskaran-marchetti2009, saintillan-shelley2013}. Since kinetic closures are typically based on assuming that the suspended particles are dilutely concentrated and interact only weakly \cite{marchetti-review}, it is of interest to develop alternative coarsening procedures which are free of these limitations. Promising alternatives to kinetic theory include, for example, coarsening procedures based on direct calculation of ensemble averages~\cite{Chuang2007} and coarsening procedures based on space-time averages. In this article, we focus on the latter.

The primary objective of this paper is to develop a coarsening method for active suspensions that does not require a kinetic formulation and that can deal with highly concentrated and strongly interacting particle systems. To focus on features induced by self-propulsion, we work with spherical particles (modeled as point particles interacting with appropriate forces) instead of rod-like particles. This makes sense for the following reasons. First, simulations show that suspensions of active point particles manifest a rich variety of semi-ordered and ordered states \cite{HPKF}, including vortical, meso-turbulent,  and polar ones. Second, it is important to understand, by means of a consistent bottom-up derivation, which features of the continuum equations are due mainly to self-propulsion and high concentration in contrast to attributes that stem from the presence of orientational degrees of freedom. 
Third, the models using spherical particles are simpler than those using elongated-particle models, which typically require additional balance equations for director variables. On this basis, it seems reasonable to develop microstructure-consistent equations for a simpler case (which is done here), and to leave more complicated cases involving, say, orientational degrees of freedom for future research. Finally, the equations developed in this paper can be directly applicable to experimental situations involving spherical swimming bacteria such as a strain of
\emph{Serratia marcescens} studied in \cite{rabani}. More generally, as Dusenbery \cite[page~25]{dusenbery} reports, about 
10\% of motile bacterial species are spherical. 


Starting from a fine-scale dissipative particle dynamics (DPD) \cite{dpd-first, espanol-warren} model of self-motile point particles, we derive meso-scale continuum equations by applying an averaging technique known as the Irving--Kirkwood--Noll procedure \cite{Kirkwood,Noll}. The most relevant version of this approach is that introduced by Hardy \cite{Hardy} and, later and independently, by Murdoch and Bedeaux \cite{mb, mb96, mb97, murdoch-book}. In this procedure, continuum equations are derived systematically and directly from particle equations.
A collision-based kinetic formulation is thus avoided. In our context, the averages depend on a mesoscopic length scale $\eta$ that is assumed to be much larger than  the range $R$ of the DPD forces.  The ratio $\eta/R$ embodies a separation of scales. In contrast to results that rely on ensemble averaging, our equations describe single realizations, both in terms of initial conditions and realizations
of fluctuation forces. This has significant practical advantages because stochastic averaging, which is commonly associated with large errors and high computational costs, is no longer required to calculate effective parameters.  Explicit constitutive equations are obtained from spatial statistics of fluctuations. In particular, our method results in the commonly encountered cubically nonlinear (internal) body force density associated with self-propulsion. Surprisingly, this arises even though the self-propulsion force in our DPD model does not incorporate a mechanism for aligning neighboring particles or other velocity selection mechanisms.

Our effective continuum equations resemble those arising in the well-known phenomenological model of Toner and Tu \cite{toner-tu95}. Toner--Tu type equations have previously been derived by applying the kinetic theory to systems of self-propelled particles that move with constant speed
in directions that change in response to
a velocity-aligning force \cite{bertin06,ihle11}. It is noteworthy that the same set of continuum equations results from a completely different microscopic model in which short-range interactions dominate but no velocity-aligning force appears.

Our continuum equations combine features of three classical models: the Navier--Stokes equations for a compressible fluid, the Ginzburg--Landau equations, and the Langevin equations.  The structure of the momentum balance and a linear constitutive equation for viscous stress are reminiscent of the Navier--Stokes equations for a compressible fluid, 
with the hydrostatic contribution to the pressure being determined as a nonlinear function of the effective mass density. The viscosity tensor is also given by a constitutive relation. The cubic nonlinearity of the effective self-propulsion forces is typical of the Ginzburg--Landau equations. A feature of our approach which brings to mind Langevin equations is the presence of an additive stochastic force.

As a consequence of the fluctuation-dissipation relation \cite{espanol-warren}, the physical temperature is incorporated into the constitutive relations via dependence of the DPD forces on the temperature. In addition, the constitutive functions depend on two scalar temperature-like parameters. While one of these characterizes fluctuations of fine scale (DPD) velocities and can be associated with an ``upscaling temperature," the other describes the extent to which relative particle positions fluctuate. For sufficiently dense and slightly compressible systems, position fluctuations tend to remain close to their initial values and, thus, they may be determined without using any fine scale computing. Averaging of random DPD forces yields a coarse-scale fluctuation force which is reminiscent of the stochastic driving term in the Langevin equation.  The force is uncorrelated in time and realizes a Gaussian random field at each instant of time. However, in contrast to the Langevin equation, the strength and the variance here are given by constitutive functions of the mass density, the temperature, and the fluctuation strength parameters of the DPD forces. Therefore, the statistics of the average fluctuation force vary with time and position. Importantly, however, the variance decreases with increasing scale separation, so that the probability distribution of the coarse-scale fluctuation force concentrates more and more tightly near the mean. Since the mean is zero, this force vanishes in the limit of infinite scale separation, and the model becomes deterministic. Suggesting that imposing a fine-scale fluctuation-dissipation relation does not in general yield an analogous coarse-scale relation, this construction may be of independent interest.

The remainder of the paper is organized as follows. The DPD model is described in Section \ref{sect:dpd}. Exact averaged equations are provided in Section \ref{sect:exact-av}. Our closure method, which is contingent on the spatial statistics arising for the discretized formulation of the balance equations, is described in Section~\ref{sect:statistics}. The resulting closed-form continuum equations are summarized in Section~\ref{sect:cont-summary}. Closed-form approximations for the  average self-propulsion force and the convective stress that models the momentum transfer due to velocity fluctuations are derived in Section~\ref{sect:sp-cs-closure}. The average fluctuation force equation is derived in Section~\ref{sect:rand1} and analyzed in Section~\ref{sect:rand2}. A linear stability analysis is conducted is Section \ref{sect:lin-stab}. Closed-form approximations for the stresses induced by conservative and dissipative forces are derived in Sections~\ref{sect:cons-closure} and Section~\ref{sect:diss-closure}, respectively. A synopsis of our most salient results is provided in Section \ref{sect:conclusions}. Closed-form approximations for the stresses induced by conservative and dissipative forces are derived in the Appendix.
\section{DPD equations of motion}
\label{sect:dpd}

Dissipative particle dynamics have been previously used for modeling of passive colloidal suspensions at low, moderate, and high concentrations \cite{dense1-boek, dense2-bolint, dense3-laurati}.  For high concentrations, DPD equations with short-range pairwise forces seem to work reasonably well because long-range hydrodynamics interactions are screened by nearly touching neighboring particles. (See, for example, \cite{panchenko-jfm} for a detailed mathematical treatment of the screening effect.) 

It therefore seems natural to develop a suitable DPD model for active colloidal suspensions. To this end, we augment the conventional DPD forces (to be detailed shortly) 
with a self-propulsion force
\begin{equation}
\label{dpd-sp-force}
\fbf_i^{\textit{SP}}=A^{\textit{SP}}h(|\bv_i|)\bv_i,
\end{equation}
where $A^{\textit{SP}}$ is a constant strength parameter with dimensions of force and $h$ is a nonnegative function with dimensions of inverse velocity.  A related observation is that the product $A^{SP}h(|\bv|)$ carries dimensions of mass per unit time or, equivalently, of viscosity per unit length. This product can therefore be viewed as a velocity-dependent viscosity coefficient. Other properties of $h$ will be discussed in the final paragraph of this section.  

To simplify the presentation, we assume that all DPD particles have equal mass $m$. The positions $\bq_i$ of a particles evolve according to the ordinary-differential equations 
\begin{equation}
m\ddot\bq_i
=\fbf_i^{\textit{SP}}+\sum_{j\ne i}\fbf^C_{ij}+\sum_{j\ne i}\fbf^D_{ij}+\sum_{j\ne i}\fbf^R_{ij},~~~~~~~i, j=1, 2, \ldots, N,
\label{dpd-odes}
\end{equation}
where $\fbf^C_{ij}$, $\fbf^D_{ij}$, and $\fbf^R_{ij}$ are the conservative, dissipative, and fluctuation forces familiar from conventional DPD \cite{dpd-first, espanol-warren}. Specifically,
\begin{align}
\fbf^C_{ij}&=A w^C(r_{ij}) \be_{ij},
\label{dpd-c-force}
\\[4pt]
\fbf^D_{ij}&=-\gamma w^D(r_{ij})(\bv_{ij}\cdot\be_{ij})\be_{ij},
\label{diss-force}
\\[4pt]
\fbf^R_{ij}&=\alpha \xi_{ij} w^R (r_{ij})\be_{ij},
\label{rand-force}
\end{align}
where $A$ is a stiffness coefficient, $\gamma$ is a drag coefficient, $\alpha$ is a strength parameter, $w^C$, $w^D$, and $w^R$ are window functions, $\xi_{ij}$ is a Gaussian random variable with zero mean and unit variance satisfying $\xi_{ij}=\xi_{ji}$, and $r_{ij}$ and $\be_{ij}$ are given by
\begin{equation}
r_{ij}=|\bq_i-\bq_j|  
\qquad\text{and}\qquad
\be_{ij}=\frac{\bq_i-\bq_j}{|\bq_i-\bq_j|}.
\end{equation}
The parameters are related by the fluctuation-dissipation relations \cite{espanol-warren}
\begin{equation}
w^D=\left(w^R \right)^2,\qquad\alpha^2=2\gamma k_B T,
\label{FDrelns}
\end{equation}
where $k_B$ is the Boltzmann constant and $T$ is the absolute temperature. These relations are the consequence of imposing a balance of fluctuation forces and dissipative forces so that the associated Fokker--Planck equation has a steady state solution given by the Gibbs canonical probability density $Z^{-1}e^{-(k_B T)^{-1} H}$, where $H$ is the sum of the kinetic energy and the potential energy of the conservative DPD forces, and $Z$ is a normalizing constant known as the partition function.

\begin{figure}[!t]
\label{Fig:sp-app}
\centering
\includegraphics[width=0.416\textwidth]{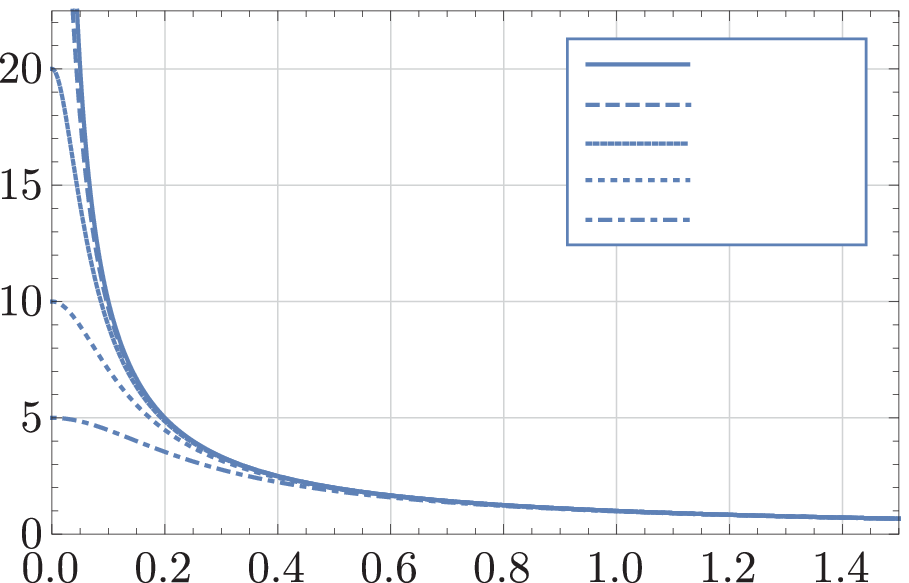}
\put(-227,60){\rotatebox{90}{$h(\xi)$}}
\put(-110,-13){$\xi$}
\put(-40,120){\tiny$1/\xi$} 
\put(-40,111){\tiny$\delta=0.025$}
\put(-40,102){\tiny$\delta=0.050$}
\put(-40,93){\tiny$\delta=0.100$}
\put(-40,84){\tiny$\delta=0.200$}
\caption{Approximation of $1/\xi$ by $h(\xi)=1/\sqrt{\xi^2+\delta^2}$ for various values of $\delta$.}
\end{figure}

From its definition \eqref{dpd-sp-force}, the self-propulsion force $\fbf_i^{\textit{SP}}$ aligns with the direction $\bv/|\bv|$ determined by the velocity vector $\bv_i$. 
%
%
One possibility is to choose $h$ such that
\begin{equation}
\label{choice-h}
h(\xi)=\frac{1}{\sqrt{\xi^2+\delta^2}}.
\end{equation}
In this case, the magnitude of the self-propulsion force $\fbf_i^{\textit{SP}}$ defined by \eqref{dpd-sp-force} is nearly equal to the constant strength parameter $A^{\textit{SP}}$ for $\delta\ll|\bv|$, in which case the orientation $\bv/|\bv|$ is closely approximated by $\bv/\sqrt{|\bv|^2+\delta^2}$.  The approximation is illustrated in Figure \ref{Fig:sp-app}.

The particular value of $\delta$ should be selected consistent with the requirement that, at any given instant of time, most particles move with the velocities that are considerably larger than $\delta$. Thus, 
$\delta$ should be small in comparison to $\sqrt{K}$, where $K$ is the spatio-temporal average kinetic energy per particle and per unit mass. Formally, choosing $\delta>0$ prevents division by zero when $\bv_i$ vanishes.  The choice \eqref{choice-h} can be viewed as a constitutive relation of stick-slip type intended to mimic the observed behavior of bacteria, which swim mostly at constant velocity but occasionally pause.
\section{Exact continuum equations. Stresses corresponding to pair forces}
\label{sect:exact-av}
To derive meso-scale continuum equations from the micro-scale model, we apply spatial averaging to a single realization of DPD equations. (Here, the term ``single realization'' refers to one realization of stochastic forces present in the DPD model.)
The initial conditions for the DPD equations are assumed to be both deterministic and precisely known. In that sense, our strategy is therefore predicated on conditions that differ significantly from those 
commonly used to justify 
ensemble averaging in statistical mechanics. The expected result of spatial averaging is a system of stochastic continuum equations. The random nature of these equations should be inherited from the underlying stochastic ordinary differential equations.

Spatial averages are defined using a window function $\psi_\eta$. This function depends on the averaging length scale $\eta$ and is normalized to have a unit integral for each $\eta$.  Most often, $\psi_\eta$ is non-negative and either compactly supported or rapidly decreasing.
The average mass density $\overline\rho$ and linear momentum $\overline{\bp}$ are defined according to \cite{Hardy}
\begin{equation}
\label{gen-density}
\overline{\rho}(t, \bx)=\sum_{j=1}^N m_j \psi_\eta (\bx-\bq_j(t))
\end{equation}
%
and
\begin{equation}
\label{gen-momentum}
\overline{\bp}(t, \bx)=\sum_{j=1}^N m_j \bv_j \psi_\eta (\bx-\bq_j(t)).
\end{equation}
Mimicking conventional derivations \cite{Hardy,mb}, we may differentiate these quantities with respect to time and then use the ordinary-differential equations of the DPD model to eliminate time derivatives of the particle velocities. This yields exact balance equations for mass and linear momentum of the form
\begin{equation}
\label{exact-density}
\partial_t \overline{\rho}+{\rm div}\overline{\bp}=0
\end{equation}
and
\begin{equation}
\label{exact-momentum}
\partial_t \overline{\bp}+{\rm div}(\overline{\bp}\otimes \overline{\bv})={\rm div}
\left[
-\sum_{j=1}^N m_j (\bv_j-\overline{\bv})\otimes (\bv_j-\overline{\bv})\psi_\eta(\bx-\bq_j)\right]
+{\rm div}\bT+{\bg}^R+ {\bg}^{\textit{SP}},
\end{equation}
where 
\begin{equation}
\label{average-velocity}
\overline{\bv}=\frac{\overline\bp}{\overline\rho}
\end{equation}
is the average velocity, $\bT$ is the interaction stress, 
\begin{equation}
\label{R0}
{\bg}^R(t, \bx)=\sum_{j, k=1}^N \fbf_{jk}^R \psi_\eta(\bx-\bq_j(t))
\end{equation}
is the average fluctuating force, and
\begin{equation}
\label{av-sp}
{\bg}^{\textit{SP}}(t, \bx)=\sum_{j=1}^N \fbf_j^{\textit{SP}}\psi_\eta(\bx-\bq_j(t))
\end{equation}
is the average self-propulsion force. 

The interaction stress $\bT$ consists of a sum
\begin{equation}
\label{exact-stress}
\bT=\bT^C+\bT^D,
\end{equation}
of conservative and dissipative contributions, where $\bT^C$ and $\bT^D$ are determined respectively by the DPD pair forces $\fbf^C_{ij}$ and $\fbf^D_{ij}$ defined in \eqref{dpd-c-force} and \eqref{diss-force} through
\begin{equation}
\label{cons-stress}
\bT^C=\frac 12 \sum_{i=1}^N \sum_{j=1}^N \fbf^C_{ij} \otimes (\bq_j-\bq_i) \Psi_\eta(\bx, \bq_i, \bq_j)
\end{equation}
and
\begin{equation}
\label{diss-stress}
\bT^D=\frac 12 \sum_{i=1}^N \sum_{j=1}^N \fbf^D_{ij} \otimes (\bq_j-\bq_i) \Psi_\eta(\bx, \bq_i, \bq_j),
\end{equation}
with $\Psi_\eta$ being defined by 
\begin{equation}
\label{functPsi}
\Psi_\eta(\bx, \bq_i, \bq_j)=\int_0^1 \psi_\eta (s(\bx-\bq_i)+(1-s)(\bx-\bq_j))\,\text{d}s.
\end{equation}
On using \eqref{dpd-c-force} and \eqref{diss-force} in \eqref{cons-stress} and \eqref{diss-stress}, $\bT^C$ and $\bT^D$ are seen to be symmetric. Thus, by \eqref{exact-stress}, the interaction stress obeys
\begin{equation}
\bT=\transpose{\bT},
\end{equation}
which is not surprising for a system of point-like particles (regardless of activity).

Although the average fluctuation force $\bg^R$ can be also written as the divergence of the random interaction stress
\begin{equation}
\label{rand-stress}
\bT^R=\frac 12  \sum_{i=1}^N \sum_{j=1}^N \fbf^R_{ij} \otimes (\bq_j-\bq_i) \Psi_\eta(\bx, \bq_i, \bq_j),
\end{equation}
working directly with $\bg^R$ is more convenient because its statistics are somewhat more easily described and because doing so results in a model that resembles the Langevin equation.

It does not seem possible to apply the simple techniques of Noll \cite{Noll} and Hardy \cite{Hardy} to express the self-propulsion force $\bg^{\textit{SP}}$ as the divergence of a stress-like object. We have therefore opted to treat this force as a body force in the continuum theory.

%
The exact equations \eqref{exact-density}--\eqref{exact-momentum} do not constitute a continuum model in the true sense. Indeed, the stress and other terms in the right hand side of \eqref{exact-momentum} cannot be determined without knowledge of the trajectories of all DPD particles. Since a true continuum model should be self-contained (or closed), exact equations should be supplemented by a closure approximation that allows the right-hand side of \eqref{exact-momentum} to be expressed as a function of available averages, namely the mass density, linear momentum, temperature, and so on. Finding a suitable closure is both the most difficult and the most important step in deriving meso-scale equations from a microscopic model.

\section{Average velocity, average deformation and fluctuation-based closure}
\label{sect:statistics}
\subsection{Averaging in the discrete setting}
In conventional continuum theories, averages are defined at each point of space and at each instant of time. However, in most situations of practical interest, the objective is to compute solutions using a discretized version of the governing equations. Discretization reduces the available information because the spatial resolution of any numerical method is inherently finite. In principle, this resolution can be refined indefinitely. In practice, refinement is, however, limited by the available computing power.  It is therefore natural to consider the situation where the smallest available resolution is in place and cannot be further reduced, in which case averages are available only at points $\xbeta$, $\beta=1,2,\dots,B$, of the computational grid. In principle, time should be also discretized. We nevertheless focus on spatial discretization and assume that the grid values are known at each instant of time. 

Discretizing a continuum description in accompanied by a loss of information which turns out to be quite severe. In the continuum setting, the fine-scale velocity can, in principle, be uniquely reconstructed from the knowledge of the average density and momentum \cite{PBG, PT, BP}. A unique reconstruction is possible for each realization of particle dynamics. Once velocities have been determined, integrating with respect to time leads to a unique recovery of positions. Uniqueness implies that fine-scale information is completely transferred to the meso-scale. In the discretized setting, uniqueness is lost and many different particle states can generate the same grid values of, say,  the average mass density and momentum. Missing information should be quantified using suitable statistics for fluctuations. The statistics developed here differ from the atomistic ensemble statistics, which originate from indeterminate initial conditions. Since the typical spatial scales in DPD are much larger than the distances between neighboring fluid molecules, it is reasonable to suppose that the DPD initial conditions are given precisely. Averaging of a single realization and measuring the averages on a finite set of points is nevertheless still accompanied by an information deficit.

\subsection{Average mass density and average velocity}

Consider a simple discretized averaging model in dimension $d$, with $d=2$ or $d=3$, assuming that the continuum length scale $\eta$ corresponds to the finest affordable resolution. Consider a cubic computational domain $\Omega$ with $d$-dimensional volume $\mathcal{V}$. Divide $\Omega$ into $B$ non-overlapping cubic averaging cells $\Cbeta$, $\beta=1,2,\dots,B$, each of side length $\eta$ and volume $\Vbeta=\mathcal{V}/B=\eta^d$. Let $\Cbeta$ be centered at $\xbeta$. Further, let $\Ibeta$ be the time-dependent index set of particles located in $\Cbeta$ and let $\nbeta(t)$ be the number of particles in $\Ibeta$ at time $t$. 
\begin{figure}[!t]
\label{Fig:ac-cells}
\centering
\includegraphics[width=0.9\textwidth]{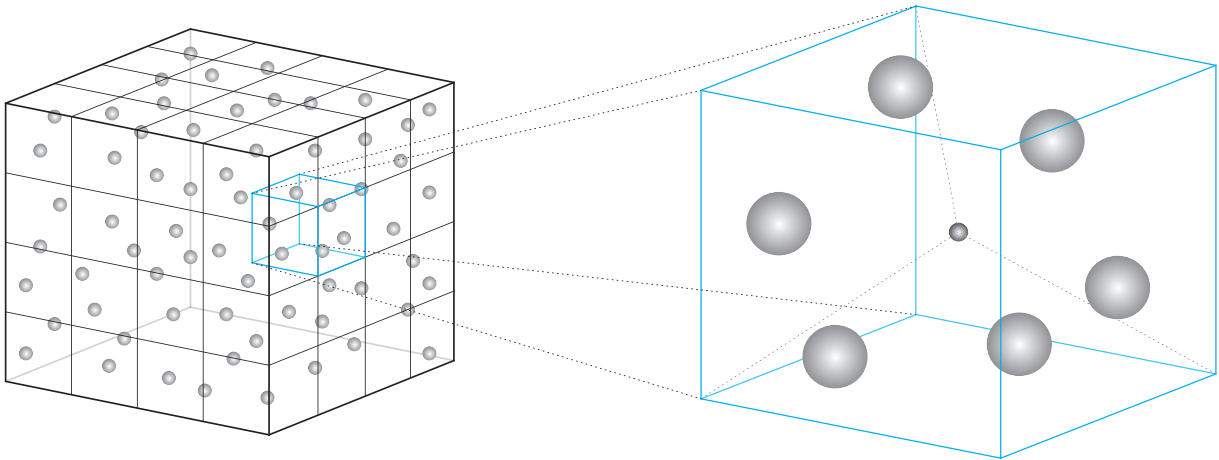}
\caption{Averaging cells within the flow domain $ \Omega$.}
\end{figure}

Define the average mass density $\overline{\rho}_\beta$ of $\Cbeta$ by
\begin{equation}
\label{av-density}
\overline{\rho}_\beta(t)=\overline{\rho}(t, \xbeta)
=\frac{1}{\Vbeta}\sum_{j\in\Ibeta}m=\frac{m\mskip1mu n_\beta(t)}{\Vbeta}
\end{equation}
where $m$ is the mass of particle $j$ in $\Ibeta$. Similarly, define the average velocity $\overline{\bv}_\beta$ of $\Cbeta$ by 
\begin{equation}
\label{v-bar}
\overline{\bv}_\beta(t)=\overline{\bv}(t,\xbeta)=\frac{1}{\nbeta}\sum_{j\in \Ibeta} \bv_j(t).
\end{equation}
%
%
Importantly, for a particle $j$ in $\Cbeta$, the velocity fluctuations 
\begin{equation}
\label{vel-fluct}
\bv_j^\prime=\bv_j-\overline{\bv}_\beta
\end{equation}
satisfy
\begin{equation}
\label{vel-zero}
\sum_{j\in J_\beta} \bv_j^\prime=0.
\end{equation}


The averages \eqref{av-density} and \eqref{v-bar} can be obtained, respectively, from \eqref{gen-density} and \eqref{gen-momentum} on choosing the window function $\psi_\eta$ to be of the particular form
\begin{equation}
\label{psi-box}
\psi_\eta(\bx)=
\left\{
\begin{array}{cc}
\displaystyle
\frac{1}\Vbeta & {\rm if}\;\bx\in \Cbeta,
\cr\noalign{\vskip4pt}
0 & {\rm otherwise}.
\end{array}
\right.
\end{equation}
Granted that all particles have equal mass and using Hardy averages \cite{Hardy}, we then have
\begin{eqnarray*}
\overline{\bv}_\beta(t)
= \frac{\sum_{j=1}^N m\bv_j \psi_\eta(\xbeta-\bq_j)}{\sum_{j=1}^N m \psi_\eta(\xbeta-\bq_j)}=
\frac{\Vbeta^{-1}\sum_{j\in \Ibeta} \bv_j }{\Vbeta^{-1}\sum_{j\in \Ibeta} 1}=
\frac{1}{\nbeta}\sum_{j\in \Ibeta} \bv_j.
\end{eqnarray*}
More generally, we may write
\begin{equation}
\label{average-velocity-gen}
\overline{\bv}(t, \bx)
=\frac{\sum_{j=1}^N m \bv_j \psi_\eta(\bx-\bq_j(t))}{\sum_{j=1}^N m \psi_\eta(\xbeta-\bq_j(t))}=
\frac{1}{n(t, \bx)}\sum_{j\in I(t, \bx)} \bv_j,
\end{equation}
where $n(t,\bx)$ is the number of particles in the cube ${\mathcal C}_\bx$ with the volume $\Vbeta$ centered at $\bx$ at time $t$, and $I(t,\bx)$ is the associated index set of particles located within this box.
%

%


\subsection{Average deformation}


By analogy to the classical kinematical connection between the referential and spatial descriptions of velocity, we define the average deformation $\bchi$ by
$$
\dot\bchi(t, \bX)=\overline{\bv}(t, \bchi(t, \bX)),\;\;\;\;\bchi(0,\bX)=\bX,
$$
where $\bX$ denotes a generic point in a fixed reference configuration. Although this quantity is not known a priori, it is useful to represent the relative particle positions 
$$
\bq_{ij}(t)=\bq_i(t)-\bq_j(t)
$$
in the form
$$
\bq_{ij}=\overline{\bq}_{ij}+\bq_{ij}^\prime,
$$
where $\overline{\bq}_{ij}=\overline{\bq}_i-\overline{\bq}_j$ is an average relative position compatible with $\bchi$, and $\bq_{ij}^\prime$ is the fluctuation. Compatibility is understood as follows. Given that $\bq_i$ and $\bq_j$ are in $\Cbeta$, the average positions are prescribed by
\begin{equation}
\label{av-positions}
\overline{\bq}_{ij}= \nabla \bchi (t, \bX_\beta)(\bX_i-\bX_j),
\end{equation}
%
where $\bX_\beta$ is a point which for present purposes is associated with the cell $\Cbeta$.  
A natural choice of $\bX_\beta$ is
$$
\bX_\beta=\bchi^{-1}(t, \xbeta)
$$
(the pre-image of the cell center under the inverse average deformation map). 
It is important to recognize that, in general, $\bX_\beta$ may change in time, and that it need not lie within $\Cbeta$.

The points $\bX_i$ and $ \bX_j$ lie on a fixed periodic lattice
which may be identified with the undeformed reference lattice covering the whole initial flow domain $\Omega$. The corresponding lattice vectors have equal length determined by placing $N$ particles at $\bX_i$ in $\Omega$.
The actual initial positions $\bq_i^0, i=1, 2,\dots,N$, constitute a perturbation of the reference lattice. These perturbations need not be small. We  require only that the distribution of the fluctuations $\bq_i^{\prime, 0}=\bq_i^0-\bX_i$ be orientation-independent in the sense to be made precise in Sect.~\ref{empiricalstatistics}.

With this choice of $\bX_i$, the points $\overline{\bq}_{i}(t)$ within each cell $\Cbeta$ also form a lattice at each instant $t$. The deformations of these lattices relative to the reference lattice are determined by $\nabla\bchi(t, \xbeta)$.  The deformed lattices may consequently differ from cell to cell. In particular, differences in lattice orientation may account for meso-scale vortices.



%
%

Once $\overline{\bq}_i$, $i=1,2,\dots,N$, are chosen, the fluctuations are determined by the actual particle positions, and we require that
\begin{equation}
\label{zero-av-fluct}
\sum_{i, j\in \Ibeta} \bq_{ij}^\prime=\bf0.
\end{equation}
This assumption is consistent with the expectation that particles can be locally ordered on the basis of the average deformation gradient, and, moreover, that the differences between actual and average positions are uncorrelated within a cell. 

In the sequel, we will analyze the forces between particles located at nearby lattice points.  
These forces depend on the average relative positions $\overline{\bq}_{ij}$, $i,j=1,2,\dots,N$.  Although the number of lattice sites within each cell is large, the assumption that the range $R$ of the DPD forces is much smaller than $\eta=\Vbeta^{1/d}$ implies that the number of relevant neighbors of any given site must be small. In any representative cell $\Cbeta$, the vector $\overline{\bq}_{ij}$ can therefore only assume one of a small number $A$ of values denoted by
\begin{equation}
\label{q-al}
\overline{\bq}^{\alpha\beta},~~~~~~~\alpha=1, 2, \dots, A,
\end{equation}
where the index $\alpha$ serves to enumerate different possible values within $\Cbeta$. 
We refer to $\overline{\bq}^{\alpha\beta}$, $\alpha=1,2,\dots,A$, $\beta=1,2,\dots,B$, as the \emph{relative lattice vectors}. The number $A$ of these vectors in a given cell $\beta$ is set by the lattice geometry, the range of the DPD forces, and the local density (average interparticle distance). For example, in a two-dimensional triangular lattice with nearest neighbor interactions, $\alpha=6$. If next-to-nearest neighbor interactions are also relevant, then $\alpha=16$. Moreover, in a three-dimensional cubic lattice with only nearest neighbors being relevant, $\alpha= 17$. Relative lattice vectors corresponding to one site of the two-dimensional triangular lattice are shown in Fig. 3.

\begin{figure}[!t]
\label{Fig:rel-lattice}
\centering
\includegraphics[width=0.375\textwidth]{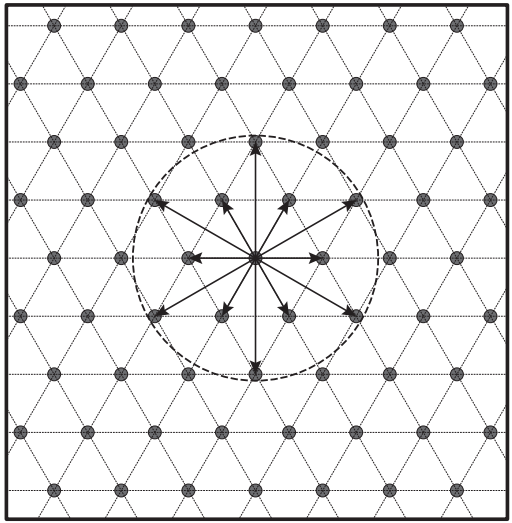}
\caption{Relative lattice vectors associated with one lattice site. The dashed circle shows the range of the DPD forces.}
\end{figure}

The extent to which local lattice vectors stretch relative to the reference lattice is determined by the parameters $s_{\alpha\beta}$, as defined by
\begin{equation}
\label{q-al1}
\left |\overline{\bq}^{\alpha\beta}\right|=s_{\alpha\beta} e_{\alpha},
\end{equation}
where $e_{\alpha}$ is the length of the corresponding local lattice vector in the underformed lattice.


When a local deformation gradient is nearly spherical, the corresponding local lattice deformation is close to a uniform expansion (or contraction), as characterized by
\begin{equation}
\label{true-stretch}
\hat s_\beta(t)=\left({\rm det}\nabla\bchi(t, \xbeta)\right)^{1/d}. 
\end{equation}
Given a particular lattice geometry, it is also possible to estimate $\hat s_\beta$ using the average concentration
\begin{equation}
\frac{\nbeta}{\Vbeta}.
\end{equation}
Indeed, the number $\nbeta$ of particles within the cell $\Cbeta$ can be approximated by
$$
\nbeta\approx \frac{\Vbeta}{c_d(\hat s_\beta{\ell})^d},
$$
where $c_d$ is a constant that depends only on the lattice geometry and the dimension $d$ of the underlying point space, and where $\ell$ is the length of the reference lattice vector connecting nearest neighbors. The concentration can be related to the mass density $\overline{\rho}_\beta=\overline{\rho}(\cdot,\xbeta)=m\mskip1mu\nbeta/\Vbeta$ 
by writing
$$
\frac{1}{(\hat s_\beta \ell)^d}\approx \frac{c_d\mskip1mu\nbeta}{\Vbeta}=\frac{c_d\mskip1mu\overline{\rho}_\beta}{m},
$$
where we recall that $m$ denotes the mass of a single DPD particle. Solving the foregoing relation for $\hat s_\beta$ yields
\begin{equation}
\label{length-density}
\hat s_\beta \approx \frac{1}{\ell}\left(\frac{m}{c_d\mskip1mu\overline{\rho}_\beta}\right)^{1/d}.
\end{equation}

Hereafter, it is convenient to utilize the local lattice length scale $|\overline{\bq}|_\beta$ defined by
\begin{equation}
\label{local-length}
|\overline{\bq}|_\beta=\hat s_\beta{\ell}.
\end{equation}


It is worth mentioning that
the  affine component  $\overline{\bq}_{ij}^\beta=\nabla\bchi_\beta (\bX_i-\bX_j)$ of the relative displacement in the cell 
${\mathcal C}_\beta$ is obtained by identifying the actual particle deformations with the average deformation $\bchi$. It can be thus said that $\overline{\bq}_{ij}$ complies with the Cauchy--Born rule, while the fluctuating component $\bq_{ij}^\prime$ describes possible violations of the rule.

While the Cauchy--Born rule is typically associated with deformations of crystalline materials, it should be still relevant for dense fluids and soft matter provided that kinetic energy of velocity fluctuations is sufficiently small compared to the kinetic energy of the average motion. 
Moreover, our averaging scheme rests on the introduction of a lattice. This is done to enable efficient calculation of the constitutive equations, since for periodic arrays, the pair contributions in the constitutive equations become highly repetitive.  This does not prohibit geometric irregularities in the actual particle placement. Rather, introduction of the local lattice corresponds to the expectation that average deformation gradient is close to piecewise constant (and, thus, that the average deformation itself is nearly piecewise linear) at the chosen meso-scale. Granted that the separation between the averaging scale and the fine scale is sufficiently large and that the initial conditions do not contain strong oscillations at the fine scale, the average deformation should be free of small-scale fluctations, and the above perturbed lattice picture of the local deformation should be reasonable. It is also worth noting that imposing \eqref{zero-av-fluct} is tantamount to stipulating that the Cauchy--Born rule holds on average. 


Lattices of many different geometries may be compatible with the same average deformation. It is therefore important to have a method for choosing a specific lattice geometry that best fits the available information, namely the values of the average mass density and velocity and the initial conditions for the DPD model. From the initial conditions we can extract the initial average coordination number in each cell. Combining this information with the knowledge of the initial mass density in this cell (which determines the particle population in each cell) provides a selection method for choosing a unique isotropic lattice at the initial instant of time. The same lattice also serves as the reference lattice. The simplest version of this approach, described above, would yield the same lattice in each cell, provided that the initial density is constant and that the coordination number is the same in all cells. This places restrictions on the initial conditions. A more sophisticated and broadly applicable variant of this approach would be to use the local cell coordination numbers together with the values of the local mass density. Such a strategy might be useful for treating non-uniform initial conditions and might cause the initial (reference) local lattices to be of different geometry---for instance, cubic in one cell and tetrahedral in another cell. 

We next provide a criterion for selecting local lattices at subsequent instants of time.  Given a cell ${\mathcal C}_\beta$, the simplest option is to assign to it the same lattice geometry chosen for the pre-image $\bchi^{-1}({\mathcal C}_\beta)$ at the initial time. 
However, doing so is not necessarily optimal because it may result in a local lattice length incompatible with the length changes induced by the average deformation. We therefore select the lattice geometry that minimizes the discrepancy between the value of $\hat s_\beta$ given by \eqref{length-density} and the value given by \eqref{true-stretch}.  
This process may result in local lattice geometry which changes in time at a given location. Of course, variations in geometry from one location to another are also possible. Overall, such an approach could be viewed as a relaxation of the standard Cauchy--Born rule. Because of its comparative flexibility, the relaxed version should be applicable to crystalline solids, amorphous solids, soft matter, and dense fluid systems at sufficiently low temperatures.


\subsection{Empirical statistics and fluctuation-based closure}
\label{empiricalstatistics}
We rely on a closure strategy that is simple in the sense that it employs Taylor approximations up to the second order in fluctuations. 
The corresponding calculations are straightforward but lengthy.  For this reason, most of the details are relegated to the Appendix. 
The resulting constitutive relations incorporate the tensorial second moments of fluctuations of both the positions and the velocities of the DPD particles. These fluctuation tensors will be now described in more detail.

The position fluctuation tensor is defined by
\begin{equation}
\label{av-fluct-tensor}
\bQ^{\prime, {\alpha\beta}}=\sum_{(i, j)\in J_{\alpha\beta}}\bq^\prime_{ij}\otimes \bq^\prime_{ij}
\end{equation}
where$J_{\alpha\beta}$ denote the index set of pairs $(i, j)$ such that $\overline{\bq}_{ij}=\overline{\bq}^{\alpha\beta}$ for some $\alpha$ within a generic cell $\Cbeta$.  
Further, velocity fluctuations are embodied by the second order tensor
\begin{equation}
\label{T0}
\overline{\bv^\prime\otimes \bv^\prime}^\beta=\frac{1}{\nbeta}\sum_{j\in \Ibeta} \bv^\prime_j \otimes \bv^\prime_j.
\end{equation}

We assume that the DPD particles in each cell are in local thermodynamic equilibrium, in which case the velocity fluctuation tensor must be nearly spherical and can be characterized by one scalar parameter through a relation of the form
%
\begin{equation}
\overline{\bv^\prime\otimes \bv^\prime}^\beta=\theta\bI,
\label{T}
\end{equation}
where $\theta$ is a temperature-like quantity describing the strength of the velocity fluctuations.


The closure method relies on the following assumptions.
\begin{enumerate}
\item \emph{The moments of all fluctuations of order greater than two are small in comparison to the second moments.}
\item \emph{The second order tensors  $\overline{\bv^\prime\otimes \bv^\prime}^\beta$ and $\bQ^{\prime, {\alpha\beta}}$ are nearly uniform in time for all combinations of $\alpha$ and $\beta$.}
\end{enumerate}
The first assumption permits us to truncate Taylor expansions to the second order in fluctuations.  Turning to the second assumption, we note that
the portion pertaining to $\overline{\bv^\prime\otimes \bv^\prime}$ is reasonable for dense, isothermal flows. A justification of the portion of assumption pertaining to $\bQ^\prime$ is provided in Appendix \ref{sect:positions-fluct}.

The second assumption also allows us to estimate fluctuations from the DPD initial conditions. Since the DPD model is already an average of a molecular model (associated with a much finer length scale), the initial conditions for DPD may be assumed to be deterministic and known precisely, as is usually done for Langevin-type equations. If the DPD initial conditions are not available, it is instead possible to specify a probability distribution of the initial conditions and to then use ensemble averaging in conjunction with spatiotemporal averaging.  

Aside from the foregoing assumptions, several additional assumptions are imposed below. These assumptions lead to significantly simplified constitutive relations. This resulting theory is physically reasonable and provides explicit constitutive relations expressing the pressure and viscosity as functions of the average deformation gradient and the fluctuation tensors entering the second of the above assumptions. If working with more complicated constitutive equations is feasible, it seems possible to relax these assumptions and develop a more accurate closed-form model on their basis.

Finally, we remark that the fluctuation-dependent quantities are reminiscent of a more general notion of ephemeral continua \cite{capriz-fried-seguin,capriz-ephemeral} From that perspective, the special nature of the case under discussion stems from the decision to explore the consequences of having only finitely many ``material points" at the mesoscopic scale. In addition to the loss of information mentioned above, the placement of the relevant points (or, equivalently, the placement of averaging cells) is related to their size, which in the present case is equal to the averaging scale $\eta$. This explicit scale dependence should be considered as one of the fundamental distinctions between a mesosopic model and a classical continuum model. An even more general framework arises when $\eta$ differs from the distance $\xi$ between the centers of two adjacent cells. The resulting constitutive equations would then depend on both length scales.  In the event that the ratio of these length scales is fixed, the features of the resulting theory are essentially the same as those of the theory presented here.  Significant differences could arise in the case when $\eta$ and $\xi$ are widely disparate, for instance when $\xi/\eta\to 0$ and at the same time
$\xi/R\to \infty$. In this case the averaging cells would significantly overlap, and a deconvolution closure strategy like that described by Panchenko, Barannyk and Gilbert \cite{PBG} could be used in conjunction with the truncated Taylor formula closure described in the subsequent sections. The purpose of the deconvolution closure would be to recover the unknown averages at the smaller scale $\xi$ from the available averages at the larger scale $\eta$. After this is done, closure could be achieved using truncated Taylor expansions and empirical fluctuation statistics.

\section{Summary of closed-form continuum equations}
\label{sect:cont-summary}

The exact equations of balance are approximated by the closed-form continuum equations
\begin{equation}
\label{mass-balance}
\partial_t \overline{\rho}+{\rm div}\overline{\bp}=0
\end{equation}
and 
\begin{equation}
\label{m-balance-closed}
\partial_t \overline{\bp}+{\rm div}(\mskip1mu\overline{\bp}\otimes \overline{\bv}\mskip1mu)=-\nabla
\left(
 \theta\mskip1mu\overline\rho
\right)
+{\rm div}{\overline{\bT}}+ (K_1(\theta)-K_2(\theta)|\overline\bv|^2) \overline\bv+\overline{\bg}^R.
\end{equation}
In \eqref{m-balance-closed}, and in the remainder of the paper, we use the superposed bar notation to emphasize the fundamental difference between an exact quantity such as the stress \eqref{exact-stress} and its closed-form approximation \eqref{total-stress-summary} in the form of a constitutive equation.

The first term on the right-hand side of \eqref{m-balance-closed} is the effective convective stress given by the gradient of the corresponding pressure $\theta\overline{\rho}$. The quantity ${\overline\bT}$ in the second term on the right-hand side of \eqref{m-balance-closed} is the effective interaction stress and will be discussed below. The third term on the right-hand side of \eqref{m-balance-closed}, which contains $\bv$, is the effective self-propulsion force. Finally, $\overline{\bg}^R$ represents a closed-from approximation of the average fluctuation force. 

Since the constitutive relation for the effective interaction stress turns out to be conventional, we include the derivation of the constitutive equations for the convective stress, self-propulsion forces, and the average fluctuation force in the main body of the paper and relegate the laborious calculations involved in the derivation of $\overline{\bT}$ to the Appendix.


Before turning to the derivations, some comments are in order. We first consider the self propulsion force and  $\overline{\bg}^R$. The constants $K_1$ and $K_2$ in the definition of the self-propulsion force obey
\begin{equation}
K_1>0
\qquad\text{and}\qquad
K_2>0.
\end{equation}
We therefore see that our method recovers the typical cubic nonlinearity which appears with only intuitive justification in many phenomenological models of collective behavior (see, for example, Toner and Tu \cite{toner-tu95}, Dunkel et al.~\cite{dunkel2013}, and Marchetti et al.~\cite{marchetti-review}). We arrive at this expression by rigorously upscaling a physically realistic microscale DPD self-propulsion force~\eqref{dpd-sp-force}. Importantly that force bears no resemblance to effective force that is obtained by upscaling. Note that the form of the continuum self-propulsion term mainly depends on how \eqref{dpd-sp-force} and \eqref{choice-h} are chosen, which we decide on the basis of the microscopic physics, as discussed in Section~\ref{sect:dpd}.

The average fluctuation force $\overline{\bg}^R$ is a Gaussian random field
with vanishing mean and variance $\sigma$. The variance is a time- and position-dependent state variable determined constitutively as a function of $\nabla\bchi$, $\theta_q$, and the temperature $T$. When the gradient of the local average deformation is nearly spherical, constitutive dependence of $\sigma$ on $\nabla\bchi$ reduces to dependence on the average mass density $\overline\rho$. 

The constitutive equation for the variance appears to be a new contribution which may be of broad interest in developing stochastic evolution equations for active continua. In the phenomenological approach, the macro-scale fluctuation force is often linked to the fluid viscosity by a formally postulated fluctuation-dissipation relation. In contrast, bearing in mind that such a relation must hold at the micro-scale (see Eq. \eqref{FDrelns}), the parameters of the fluctuation force at the macro-scale cannot be chosen based on the fluctuation-dissipation relation. Instead, the variance of the fluctuation force is expected to vary in space and time following the evolution of local particle patterns. Put differently, the more concentrated an active suspension is, and the greater the tendency of the system to self-organize, the less likely it is for the average fluctuation force to exhibit a fixed variance. Therefore, one of the more important contributions of this work is the quantification of this expectation in the form of a constitutive relation for $\sigma$. That relation appears to yield a useful refinement of the Toner--Tu equations, which do not posses this feature.  Further testing of this finding by simulation and experiments may be warranted.

Another important comment concerning $\overline{\bg}^R$ is that the variance $\sigma$ depends on the extent to which the scales are separated, as dictated by the ratio of the cell size $\eta$ to the typical range of $R$ of the DPD forces. In Section \ref{sect:rand2}, we prove that $\sigma\to0$  as $\eta/R\to 0$. Consequently, we infer that the model becomes deterministic in the limit of infinite scale separation.


Finally, we comment briefly on the constitutive relation 
\begin{equation}
\label{total-stress-summary}
{\overline{\bT}}=-P(\overline{\rho}, \theta_q) \bI+\bmu(\overline{\rho}, \theta_q) \be(\mskip1mu\overline\bv\mskip1mu)
\end{equation}
for the interaction stress. In \eqref{total-stress-summary}, $\be(\mskip1mu\overline\bv\mskip1mu)$ denotes the symmetric component of the gradient of the average velocity.  The overall structure of this constitutive relation is therefore reminiscent of that underlying the Navier--Stokes equations. That said, we emphasize that the pressure $P$ is determined by an unconventional equation of state in terms of the average mass density $\overline{\rho}$ and the fluctuation strength $\theta_q$ of relative positions. Importantly, $\theta_q$ is generally distinct from the fluctuation strength $\theta$ of relative velocities. Like $P$, the viscosity tensor $\bmu$ generally varies with both $\overline\rho$ and $\theta_q$.
%
%
Derivations of the conservative and viscous contributions to the interaction stress $\overline{\bT}$ 
are provided in Sections~\ref{sect:cons-closure} and \ref{sect:diss-closure} of the Appendix.

\section{Averaging the self-propulsion force and convective stress}
\label{sect:sp-cs-closure}
\subsection{Self-propulsion force}
For $\psi_\eta$ of the form \eqref{psi-box},  the average self-propulsion force density (which is a nonlinear volume average) is given by
$$
{\bg}_\beta^{\textit{SP}}=\frac{1}{\nbeta} \sum_{j\in \Ibeta} g(|\bv_j|^2) \bv_j=\sum_{j\in \Ibeta} h(|\bv_j|) \bv_j.
$$
We now approximate $\bg_\beta^{\textit{SP}}$ by a function of the average velocity $\overline{\bv}_\beta$. 
Toward this, we write
$$
\bv_j=\overline{\bv}_\beta+\bv^\prime_j,
$$
and use Taylor's theorem to expand $g(|\overline{\bv}_\beta+\bv^\prime_j|^2)(\overline{\bv}_\beta+\bv^\prime_j)$ about $\overline{\bv}_\beta$. Keeping only terms up to the second order in the velocity fluctuation $\bv^\prime_j$, we find that
\begin{multline}
g(|\overline{\bv}_\beta+\bv^\prime_j|^2)(\overline{\bv}_\beta+\bv^\prime_j)
=
g(|\overline{\bv}_\beta|^2)\overline{\bv}_\beta+2g^\prime(|\overline{\bv}_\beta|^2) \overline{\bv}_\beta \cdot \bv^\prime_j  \overline{\bv}_\beta +g(|\overline{\bv}_\beta|^2) \bv^\prime_j
\\
+
2 g^{\prime\prime}(|\overline{\bv}_\beta|^2)\overline{\bv}_\beta 
(\overline{\bv}_{\beta}\otimes \overline{\bv}_\beta ) :  
({\bv}^\prime_j \otimes {\bv}^\prime_j) \overline{\bv}_\beta\\
+ 
g^\prime (|\overline{\bv}_\beta|^2) |\bv^\prime_j|^2 \left({\bv}^\prime_j \otimes {\bv}^\prime_j \right)  \overline{\bv}_\beta
+
2 g^\prime (|\overline{\bv}_\beta|^2) \overline{\bv}_\beta \cdot \bv^\prime_j \bv^\prime_j+ \cdots.
\label{gex1}
\end{multline}
Averaging both sides of \eqref{gex1} while taking into consideration the identity $\nbeta^{-1} \sum_{j\in \Ibeta} \bv^\prime_j=0$ together with \eqref{T}, we obtain
\begin{equation}
\label{v-av}
\frac{1}{\nbeta} \sum_{j\in \Ibeta}
g(|\overline{\bv}_\beta+\bv^\prime_j|^2)(\overline{\bv}_\beta+\bv^\prime_j) \approx
(g(|\overline{\bv}_\beta|^2)+ (2+d)\theta g^\prime(|\overline{\bv}_\beta|^2)+
2\theta g^{\prime\prime}(|\overline{\bv}_\beta|^2)|\overline{\bv}_\beta|^2)
 \overline{\bv}_\beta,
\end{equation}
where, as before, $d$ is the spatial dimension.

For small $|\overline{\bv}_\beta|$, \eqref{v-av} simplifies further to
$$
g(|\overline{\bv}_\beta|^2)+ (2+d)\theta g^\prime(|\overline{\bv}_\beta|^2)+
2\theta g^{\prime\prime}(|\overline{\bv}_\beta|^2)|\overline{\bv}_\beta|^2 \approx
K_1 - K_2 |\overline{\bv}_\beta|^2,
$$
where $K_1$ and $K_2$ are defined by
\begin{equation}
K_1= g(0)+ (2+d)\theta g^\prime(0)
\qquad\text{and}\qquad
K_2 = -(4+d)\theta g^{\prime\prime}(0).
\label{K12gen}
\end{equation}

For the particular choice $g(\xi)=(\xi^2+\delta^2)^{-1/2}$, \eqref{K12gen} specializes to yield
\begin{equation}
\label{Ks}
K_1=g(0)=\delta^{-1}
\qquad\text{and}\qquad
K_2=(4+d)\theta \delta^{-3},
\end{equation}
which results in the constitutive equation
\begin{equation}
\label{self-prop-const}
{\bg}^{\textit{SP}}\approx \overline{\bg}^{\textit{SP}}=
(\delta^{-1}-(4+d)\theta \delta^{-3} |\overline{\bv}_\beta|^2) \overline{\bv}_\beta.
\end{equation}

\subsection{Convective stress}
With reference to \eqref{exact-momentum},  the convective stress is given by
\begin{equation}
m\sum_{i=1}^N\bv^\prime_i\otimes \bv^\prime_i \psi_\eta(\bx-\bq_i).
\label{convstress}
\end{equation}
If the window function $\psi_\eta$ has the particular form \eqref{psi-box} and the velocity fluctuations obey the assumption \eqref{T}, then \eqref{convstress} specializes to
\begin{equation}
\label{conv-closure}
\frac{m}{\Vbeta} \sum_{i\in \Ibeta} \bv^\prime_i\otimes \bv^\prime_i =
\frac{m}{\Vbeta} \nbeta \theta \bI=\overline{\rho}_\beta\theta \bI.
\end{equation}

\section{Constitutive equation for the fluctuation force}
\label{sect:rand1}

For the particular choice \eqref{psi-box} of the weight function $\psi_\eta$, the average \eqref{R0} of the fluctuation forces $\fbf^{R}_{ij}$ defined in \eqref{rand-force} yields
\begin{equation}
\label{R1}
\bg^R(t, \xbeta)=\sum_{i, j=1}^N \fbf^R_{ij} \psi_\eta(\bx-\bq_i)=
\frac{2 \gamma k_B T}{\mathcal{V}_\beta}\sum_{i\in \Ibeta}\sum_{j=1}^N\xi_{ij}
\sqrt{w^D(r_{ij})}\,\be_{ij},
\end{equation}
where $i$ is an element of $\Ibeta$ if and only if particle $i$ is located in cell $\Cbeta$ and where we have invoked the fluctuation-dissipation relations \eqref{FDrelns}.  Nontrivial contributions to the double sum may arise under a variety of circumstances. Consider two particles labeled $i$ and $j$. Then, the corresponding contribution to \eqref{R1} is potentially nontrivial if particles $i$ and $j$ both belong to $\Cbeta$ or if particle $i$ is in  $\Cbeta$ and particle $j$ is outside of $\Cbeta$ but within the range of the force, in which case $w^D(r_{ij})\ne 0$. However, because $\fbf^R_{ij}=-\fbf^R_{ji}$, the contribution to \eqref{R1} vanishes if particles $i$ and $j$ both belong to $\Cbeta$. Thus, \eqref{R1} reduces to
\begin{equation}
\label{R2}
\bg^R(t, \xbeta)=\frac{2 \gamma k_B T}{\mathcal{V}_\beta}\sum_{i\in \Ibeta}\sum_{j\notin \Ibeta}
\xi_{ij}\sqrt{w^D(r_{ij})}\,\be_{ij}.
\end{equation}
%

Since $\bq_i$ and $\bq_j$ are random variables that depend on the history of the motion, it is quite difficult to describe the probability distribution of $\bg^R$. However, a reasonable approximation can be developed by assuming that the dynamics are discrete in time. Calculating positions and velocities at a generic time step then proceeds by (i) inserting $r_{ij}$ and $\be_{ij}$ obtained at the previous time step into the equations \eqref{rand-force}, (ii) multiplying by $\xi_{ij}$, and (iii) updating positions and velocities. The central point is that $\xi_{ij}$ are independent, identically distributed normal random variables with zero mean and unit variance, 
and $\xi_{ij}$ are produced using, for example, a suitable random number generator, and without taking into account any information about $r_{ij}$ and $\be_{ij}$. Thus $\xi_{ij}$, $i,j,=1,2,\dots,N$ may be assumed to be statistically 
independent of $r_{ij}$ and $\be_{ij}$, $i,j,=1,2,\dots,N$.
Therefore, at each time step, $\bg^R_\beta$ can be treated as a linear combination $\sum_{i \in \Ibeta} \sum_{j\notin \Ibeta}\ba_{ij} \xi_{ij}$ of normal random variables $\xi_{ij}$ with the coefficients
\begin{equation}
\ba_{ij}=\frac{2 \gamma k_B T}{\mathcal{V}_\beta}\sqrt{w^D(r_{ij})}\,\be_{ij}.
\end{equation}
Standard results from probability theory lead to the conclusion that each of the $d$ components of $\bg^R(\cdot,\bx_\beta)$ is a normal random variable with zero mean and variance

\begin{equation}
\label{variance}
\sigma^{(k)}_\beta=\frac{2\gamma k_B T}{\mathcal{V}_\beta}
\sqrt{\sum_{i \in \Ibeta}
\sum_{j\notin \Ibeta}  
w^D(r_{ij})e_{ij}^{(k)}},
\qquad 
k=1,\dots,d,
\end{equation}
where $e_{ij}^{(k)}$ is the component of $\be_{ij}$ in the direction of the $k$-th basis element. Assuming that all components of $\bg^R(\cdot,\bx_\beta)$ are equally distributed yields
\begin{equation}
\label{iso-var}
\sigma^{(k)}_\beta=\sqrt{\frac{1}{d}\sum\limits_{l=1}^d(\sigma^{(l))}_\beta)^2}=
\sigma_\beta=\frac{2\gamma k_B T}{\mathcal{V}_\beta}\sqrt{\frac{1}{d}\sum_{i \in \Ibeta}\sum_{j\notin \Ibeta}w^D(r_{ij})}\,.
\end{equation}
The variance therefore becomes another state variable that requires closure. Expanding $w^D$ to the second order in $\bq^\prime$, we find that
\begin{multline}
\label{single-term}
w^D(r_{ij})=w^D(|\overline{\bq}_{ij}|)+\left(w^D\right)^\prime(|\overline{\bq}_{ij}|)
\frac{\overline{\bq}_{ij}}{|\overline{\bq}_{ij}|}
\cdot \bq^\prime_{ij}
\\[4pt]+
\frac 12 \left[ 
\left(w^D\right)^{\prime\prime}(|\overline{\bq}_{ij}|)\frac{\overline{\bq}_{ij}\otimes \overline{\bq}_{ij}}
{|\overline{\bq}_{ij}|^2}
- \left(w^D\right)^{\prime}(|\overline{\bq}_{ij}|)\frac{|\overline{\bq}_{ij}|^2 \bI-\overline{\bq}_{ij}\otimes \overline{\bq}_{ij}}
{|\overline{\bq}_{ij}|^3}
\right]: (\bq^\prime_{ij}\otimes \bq^\prime_{ij})+ \cdots,
\end{multline}
where $\left(w^D\right)^\prime$ and denotes the derivative of $w^D$ with respect to its argument, and similarly for $\left(w^D\right)^{\prime\prime}$. 

Inserting \eqref{single-term} in \eqref{R2} and, as before, first summing the terms with $\overline{\bq}_{ij}=\overline{\bq}^{\alpha\beta}$ with a fixed relative lattice vector $\overline{\bq}^{\alpha\beta}$ and then summing over $\alpha$, we find that
\begin{multline}
\label{rand-cl1}
\sigma_\beta^2 \approx \frac{4 \gamma^2 (k_B T)^2}{d (\mathcal{V}_\beta)^2}
\left(
\sum_{\alpha} w^D(|\overline{\bq}^{\alpha\beta}|) 
|S_\alpha|
+
\sum_{\alpha} \left(w^D\right)^\prime (|\overline{\bq}^{\alpha\beta}|) 
\frac{\overline{\bq}^{\alpha\beta}}{|\overline{\bq}^{\alpha\beta}|}
\cdot \sum_{(i,j)\in S_\alpha} \bq^\prime_{ij}
\right)
\\[6pt]
+ \frac{2 \gamma^2 (k_B T)^2}{d (\mathcal{V}_\beta)^2}
\sum_{\alpha}
 \left[ 
\left(w^D\right)^{\prime\prime}(|\overline{\bq}^{\alpha\beta}|)\frac{\overline{\bq}^{\alpha\beta}\otimes \overline{\bq}^{\alpha\beta}}
{|\overline{\bq}^{\alpha\beta}|^2}\right.
- \left.\left(w^D\right)^{\prime}(|\overline{\bq}^{\alpha\beta}|)\frac{|\overline{\bq}^{\alpha\beta}|^2 \bI-\overline{\bq}^{\alpha\beta}\otimes \overline{\bq}^{\alpha\beta}}
{|\overline{\bq}^{\alpha\beta}|^3}
\right]:\sum_{(i,j)\in S_\alpha} \bq^\prime_{ij}\otimes \bq^\prime_{ij},
\end{multline}
where $S_\alpha$ is the index set defined by
\begin{equation}
S_\alpha=
\{ 
(i, j): i\in \Ibeta, j \notin \Ibeta, \overline{\bq}_{ij}=\overline{\bq}^{\alpha\beta}
\}
\end{equation}
and $|S_\alpha|$ denotes the number of elements in $S_\alpha$. 

Since $|S_\alpha|$ is expected to be large, the simplest reasonable closure assumptions are
\begin{equation}
\label{rand-cl2}
\sum_{(i,j)\in S_\alpha} \bq^\prime_{ij}={\bf 0}
\qquad{\rm and}\qquad
\sum_{(i,j)\in S_\alpha}{\bq}_{ij}^\prime \otimes {\bq}^\prime_{ij}=\theta_q |\overline{\bq}^{\alpha\beta}|^2\bI,
\end{equation}
which, when inserted into \eqref{rand-cl1}, gives 
\begin{equation}
\label{rand-cl3}
\sigma_\beta^2 \approx \frac{4 \gamma^2 (k_B T)^2}{d (\mathcal{V}_\beta)^2}
\sum_{\alpha} w^D(|\overline{\bq}^{\alpha\beta}|) 
|S_\alpha|
+
\theta_q \frac{2\gamma^2 (k_B T)^2}{d (\mathcal{V}_\beta)^2}
\sum_{\alpha}
 \left[ 
\left(w^D\right)^{\prime\prime}(|\overline{\bq}^{\alpha\beta}|)|\overline{\bq}^{\alpha\beta} |^2
- (d-1)\left(w^D\right)^{\prime}(|\overline{\bq}^{\alpha\beta}|)
{|\overline{\bq}^{\alpha\beta}|}
\right],
\end{equation}
where the trivial identity ${\rm tr}(\overline{\bq}^{\alpha\beta}\otimes \overline{\bq}^{\alpha\beta})=|\overline{\bq}^{\alpha\beta}|^2$ has been used. 
Since $w^D(|\overline{\bq}^{\alpha\beta}|)> 0$, the right-hand side of \eqref{rand-cl3} is guaranteed to be positive if $\theta_q$ is sufficiently small. More detailed analyses are possible for particular choices of $w^D$. 

Since $\overline{\bq}^{\alpha\beta}=\nabla \bchi(\xbeta)\be^{\alpha}$, where $\be^{\alpha}$ is a relative lattice vector of the undeformed reference lattice, (\ref{rand-cl3}) provides a constitutive equation for the variance, given as a function of the average deformation gradient and other material parameters such as $\theta_q$, $T$,  $\gamma$, and $w^D$.

If the local deformation is close to a uniform expansion or contraction, then $|\overline{\bq}^{\alpha\beta}|\approx 
|\overline{\bq}|^\beta l_\alpha$, where $|\overline{\bq}|^\beta$ is the length scale
of a uniformly deformed local lattice vector, as defined in \eqref{local-length}, and $l_\alpha$ is a non-dimensional parameter independent of the deformation. Since, consistent with \eqref{length-density}, $|\overline{\bq}^\beta|\sim (\overline{\rho})^{-1/d}$, the right-hand side of \eqref{rand-cl3} becomes a function of mass density but also depends on the lattice geometry, $\Vbeta$, $d$, $\theta_q$, $\gamma$, and $T$.  Thus, for nearly sperhical local deformations,
\begin{equation}
\label{sigma-iso}
\sigma_\beta^2 \approx \frac{4 \gamma^2 (k_B T)^2}{d (\mathcal{V}_\beta)^2}\left(F_1(\overline{\rho}_\beta)+
\frac 12 \theta_q F_2(\overline{\rho}_\beta)
\right),
\end{equation}
where $F_1$ and $F_2$ are determined by the following sums:
\begin{equation}
\left.
\begin{split}
\label{sigma-iso1}
F_1(\overline{\rho}_\beta) &=
\sum_{\alpha} w^D
\left(
\left(
\frac{m}{c_d \overline{\rho}_\beta}
\right)^{1/d} 
\right)
|S_\alpha|,
\\[6pt]
F_2(\overline{\rho}_\beta) &=
\sum_{\alpha}
 \left[ 
\left(w^D\right)^{\prime\prime}
\left(
(\frac{m}{c_d \overline{\rho}_\beta}
)^{1/d}
\right)
\left(\frac{m}{c_d \overline{\rho}_\beta}
\right)
^{2/d}
- (d-1)\left(w^D\right)^{\prime}
\left(
(\frac{m}{c_d \overline{\rho}_\beta}
)^{1/d}
\right)
\left(
\frac{m}{c_d \overline{\rho}_\beta}
\right)^{1/d}
\right].
%
\end{split}
\,\right\}
\end{equation}

In summary, the constitutive approximation $\overline{\bg}^R$ of the exact fluctuation force $\bg^R$ is obtained by choosing, at each instant of time,
a Gaussian random field with mean zero and variance given by \eqref{sigma-iso}. The time-correlation properties of $\overline{\bq}^R$ are identical to those of $\bg^R$.\

\section{Vanishing of the variance with increasing scale separation}
\label{sect:rand2}
Since the variance $\sigma_\beta$ is influenced by the local state of deformation, the distribution of the average fluctuation force may differ from location to location and may also evolve in time. It is therefore useful to provide as much insight as possible regarding the behavior of this distribution. With this objective in mind, we consider the impact of increasing the mesoscopic length scale $\eta$ with the range $R$ of the DPD forces held fixed. Since this corresponds to increasing the number of particles that contribute to the variance, intuition suggests that the ``randomness'' of the average fluctuation force
$\bg^R$ should decrease as $\eta$ increases. In support of this heuristic expectation, we next show that
\begin{equation}
\sigma_\beta\to0
\qquad\text{as}\qquad
\frac{\eta}{R}\to\infty. 
\label{siglim}
\end{equation}
In taking the foregoing limit, we identify $R$ with the range of $\fbf_{ij}^R$ defined, with reference to the fluctuation-dissipation relations \eqref{FDrelns}, by the support of the window function $w^D$ associated with the dissipative force $\fbf^D_{ij}$. This limiting process also requires a condition on the mass density, namely that there exists a positive number $M$, with dimensions of mass per unit volume, independent of $\eta$ and $\beta$ such that
\begin{equation}
\label{rho-upper}
\overline{\rho}_\beta\leq M.
\end{equation}

Our estimate of the variance involves two steps.
\begin{enumerate}
\item \emph{Estimate $|S_\alpha|$.} We begin with the observation that the nonvanishing contributions to $\sigma_\beta$ are comprised only of pairs $(i,j)$ such that particle $i$ lies in 
$\Cbeta$ and particle $j$ lies outside of $\Cbeta$. In addition, $\fbf_{ij}^R$ should not vanish identically, that is,  the distance $r_{ij}$ between particles $i$ and $j$ should be less than the range $R$ of the DPD forces. All such particle pairs should be located in the rectangular shell 
\begin{equation}
\label{S-bet}
S_\beta=\{ \bx\in {\bf R}^d: {\rm dist}(\bx,\partial\Cbeta)< R\}
\end{equation}
containing all points with Euclidean distance to the boundary $\partial\Cbeta$ of $\Cbeta$ less than $R$. Bearing in mind that $S_\beta$ has volume
\begin{equation}
\label{volume-S}
{\mathcal V}_{S_\beta}=(\eta+R)^d-(\eta-R)^d=2R\sum_{k=0}^{d-1}
(\eta+R)^{d-1-k}(\eta-R)^k
=2R d \eta^{d-1}+O(\eta^{d-2}),
\end{equation}
the number of particles inside $S_\beta$ can be estimated by the number density
\begin {equation}
\label{part-number}
n_{S_\beta}\approx \frac{\overline{\rho}_\beta{\mathcal V}_{S_\beta}}{m},
\end{equation}
where $m$ is the mass of one particle. Increasing $M$, if necessary, we find that
$$
n_{S_\beta}\leq \frac{2M}{m}R d \eta^{d-1}
$$
for all sufficiently large values of $\eta/R$. Thus, fixing $\alpha$ and noting that for each particle $i$ in $S_\beta$ there is at most one $\overline{\bq}_{ij}$ with
$\overline{\bq}_{ij}=\overline{\bq}^{\alpha\beta}$, we arrive at the intermediate estimate
\begin{equation}
\label{e1}
|S_\alpha| \leq \frac{2MRd\eta^{d-1}}{m}.
\end{equation}
\item \emph{Estimate the remaining terms in $\sigma_\eta$}. Since $|\overline{\bq}^{\alpha\beta}|\leq R$ and since $w^D$ is bounded along with its first and second derivatives, the sums appearing in \eqref{rand-cl3} can be estimated by
\begin{equation}
\label{e2}
\sum_\alpha (w^D)^\prime (|\overline{\bq}^{\alpha\beta}|)|\overline{\bq}^{\alpha\beta}|\leq
N_{\alpha\beta} \left(\sup \left|(w^D)^\prime\right|\right) R,
\end{equation}
where $N_{\alpha\beta}$ is the number of relevant $\overline{\bq}^{\alpha\beta}$. Importantly, this number depends on $R$ and $\overline{\rho}_\beta$. For larger values of $\overline{\rho}_\beta$, the interparticle distance decreases, whereby $N_{\alpha\beta}$ tends to increase. However, if $\overline{\rho}_\beta$ satisfies \eqref{rho-upper}, it follows that there exists a positive number
$
N_{\max}
$
depending only on $R$ and $M$ such that
$$
N_{\alpha\beta}\leq N_{\max}.
$$
Thus,
\begin{equation}
\label{e2-1}
\sum_\alpha (w^D)^\prime (|\overline{\bq}^{\alpha\beta}|)|\overline{\bq}^{\alpha\beta}|\leq
N_{\max} \left(\sup \left|(w^D)^\prime\right|\right) R
\end{equation}
and the remaining sums in \eqref{rand-cl3} can be estimated similarly.  Finally, using \eqref{e1} and \eqref{e2-1} in \eqref{rand-cl3} and recalling that $\Vbeta=\eta^d$, we obtain
\begin{multline}
\label{e3}
\sigma_\beta^2\leq \frac{8M\gamma^2 (k_B T)^2}{m}
(\sup w^D)
N_{\max} R  \eta^{-d-1}
\\
+
\theta_q \frac{2\gamma^2 (k_B T)^2}{d}
N_{\max}
 \left[ 
\sup\left|\left(w^D\right)^{\prime\prime}\right|
R^2\eta^{-2d}
+(d-1)
\sup\left|\left(w^D\right)^{\prime}\right|
R\eta^{-2d}
\right]
\\[4pt]
=
C_1(M, \gamma, T, R) 
\left(
\frac{R}{\eta}
\right)^{d+1}+
C_2(M, \gamma, T, R, \theta_q) 
\left(
\frac{R}{\eta}
\right)^{2d}
+
C_3(M, \gamma, T, R, \theta_q) 
\left(
\frac{R}{\eta}
\right)^{2d},
\end{multline}
from which we conclude that $\sigma_\beta$ obeys the limit \eqref{siglim}.
\end{enumerate}

\section{Linear stability}
\label{sect:lin-stab}

We now study linear stability of constant solutions assuming infinite scale separation, so that, consistent with \eqref{siglim}, $\bg^R=\bf0$. For simplicity, we restrict attention to two spatial dimensions, and assume that all relevant effective material parameters such the elasticity and viscosity tensors are constant and isotropic.

\subsection{Stability of uniform polar solution} 
 
We first investigate the linear stability of the uniform polar solution $\rho(t, \bx)=\rho_0$, $\bv(t, \bx)=\bv_0$, with $\rho_0$ and $\bv_0$ being constants and with $|\bv_0|=\sqrt{K_1/K_2}$.

Assuming that the mass density and velocity admit expansions of the form $\rho=\rho_0+\delta\rho$ and $\bv =\bv_0+\bvep$, with $\delta\rho\ll\rho_0$ and $|\bvep|\ll|\bv_0|=\sqrt{K_1/K_2}$, we first formally linearize the mass balance \eqref{mass-balance} to yield
\begin{equation}
\label{lin-density}
\partial_t \delta\rho+\rho_0 {\rm div}\bvep+\nabla(\delta\rho)\cdot \bv_0=0.
\end{equation}
Since $\bv_0$ is constant, the characteristic streamlines of the hyperbolic equation \eqref{lin-density} are easily determined. With this information, we find that $\delta\rho$ is given in terms of $\bvep$ by
\begin{equation}
\label{deltarho}
\delta\rho(t, \bx)=-\rho_0 \int_0^t {\rm div}\bvep(\tau,\bx+(\tau-t)\bv_0)\,\text{d}\tau.
\end{equation}
Next, we formally linearize the momentum balance \eqref{m-balance-closed} to yield
\begin{equation*}
\rho_0 (\partial_t +\bv_0\cdot\nabla) \bvep=-(P^\prime(\rho_0)+\theta)\nabla(\delta\rho)- 2K_2 (\bvep\cdot\bv_0)\bv_0+\mu(\rho_0) \Delta \bvep.
\end{equation*} 
%

Taking time-derivative of both sides, using (\ref{deltarho}) to express $\partial_t \delta\rho$, and neglecting in that expression the term containing the third derivatives of $\bvep$, we find that
\begin{equation}
\label{linmom2}
\rho_0 \partial^2_{tt}\bvep +\rho_0\bv_0\cdot\nabla\partial_t \bvep=-L\nabla{\rm div}\bvep-2 K_2(\partial_t \bvep\cdot\bv_0)\bv_0+\mu \Delta \partial_t \bvep,
\end{equation}
where we have introduced $L=\rho_0 (P^\prime(\rho_0)+\theta)$. Inserting 
\begin{equation}
\bvep=\bA e^{\sigma t-i\bk\cdot\bx}
\end{equation}
into \eqref{linmom2} yields
\begin{equation}
\label{linmom3}
\lambda \bA=M\bA
\end{equation}
where, on introducing a positively oriented Cartesian basis $\{\bim_1,\bim_2\}$ and writing $k_r=\bk\cdot\bim_r$ and $v_r=\bv_0\cdot\bim_r$, $\lambda$ and $M$ are given by
\begin{equation}
\lambda=\rho_0 \sigma^2  +\rho_0\sigma (-i\bk )\cdot \bv_0+\sigma \mu |\bk|^2
\end{equation}
and 
\begin{equation}
M=
\left(
\begin{array}{cc}
L k_1^2-2K_2\sigma v_1^2 &  Lk_1k_2-2K_2\sigma v_1 v_2\\
Lk_1k_2- 2K_2\sigma v_1 v_2 &  L k_2^2-2K_2\sigma v_2^2\\
\end{array}
\right).
\end{equation}
The characteristic equation for the matrix $M$ is
\begin{equation}
\lambda^2-({\rm tr}M)\lambda +{\rm det}M=0. 
\end{equation}
Since only long wave lengths (small $\bk$) are of interest at the meso-scale, we observe that, as $\bk\to\bf0$, ${\rm tr} M=O(1)$ and ${\rm det}M=O(|k|^2)$. Thus, making the approximation
\begin{equation}
\label{quad-app}
\sqrt{({\rm tr}M)^2-4{\rm det}M}\approx {\rm tr}M- 2\frac{{\rm det}M}{{\rm tr}M},
\end{equation}
we find two solutions,
\begin{equation}
\lambda_1={\rm tr}M-  \frac{{\rm det}M}{{\rm tr}M}
\qquad\text{and}\qquad
\lambda_2=\frac{{\rm det}M}{{\rm tr}M}.
\end{equation}
These solutions depend on both $\sigma$ and $\bk$. The possible dispersion relations $\sigma=\sigma(\bk)$ should satisfy 
\begin{equation}
\label{quad1}
\rho_0 \sigma^2  +\rho_0 \sigma (-i\bk )\cdot \bv_0+\sigma \mu |\bk|^2=\lambda (\sigma, \bk),
\end{equation}
with right-hand side being either $\lambda_1$ or $\lambda_2$. Although it is possible to solve \eqref{quad1} in closed form without approximation, the result of doing so is difficult to interpret. For this reason, we approximate $\lambda_1$ and $\lambda_2$ using Taylor's formula and keeping terms up to order $O(|\bk|^2)$. This yields
\begin{equation}
\label{two-lam}
\lambda_1\approx -2K_2 \sigma |\bv_0|^2+L|\bk|^2-\frac{L}{|\bv_0|^2} E(\bk)
\qquad\text{and}\qquad
\lambda_2\approx \frac{L}{|\bv_0|^2} E(\bk),
\end{equation}
where $E$ is given by
\begin{equation}
E(\bk)=k_1^2 v_2^2 +k_2^2 v_1^2-2k_1k_2 v_1v_2.
\end{equation}
Substituting $\lambda_1$ from (\ref{two-lam}) into (\ref{quad1}) and approximating the square root in the quadratic formula as before we obtain relations
\begin{equation}
\sigma_{1}(\bk)=i\bk\cdot\bv_0-\frac{2K_2|\bv_0|^2}{\rho_0}-\frac{\mu}{\rho_0}|\bk|^2-\frac{L}{2K_2 |\bv_0|^2}
\left(|\bk|^2-\frac{E(\bk)}{|\bv_0|^2}\right)
\end{equation}
and
\begin{equation}
\sigma_{2}(\bk)=\frac{L}{2K_2|\bv_0|^2}
\left(|\bk|^2-\frac{E(\bk)}{|\bv_0|^2}\right).
\end{equation}
To analyze the stability of the modes corresponding to $\sigma_1$ and $\sigma_2$, we can assume (without loss of generality) that $\bv_0=v_1\bim_1$. Then
\begin{equation}
\label{F}
|\bk|^2-\frac{F(\bk)}{|\bv_0|^2}=|\bk|^2-k_2^2=k_1^2\geq 0.
\end{equation}
This shows that $\sigma_1$ corresponds to a stable mode and $\sigma_2$-mode is unstable.  It is interesting to note that in the incompressible case with $\mu>0$ there is no unstable mode \cite{dunkel2013}. In contrast to incompressible models, solutions of compressible equations may therefore exhibit various vortical patterns while approaching the uniform polar (flocking) state. 

 
%

Finally, the same calculations with $\lambda_2$ in place of $\lambda_1$ yield
\begin{equation}
\sigma_3(\bk)=i\bk\cdot\bv_0-\frac{\mu|\bk|^2}{\rho_0}
\qquad\text{and}\qquad
\sigma_4(\bk)=0.
\end{equation}
The mode corresponding to $\sigma_3$ is stable and the mode corresponding to $\sigma_4$ is neutrally stable.

\subsection{Stability of the trivial solution}

Formal linearization of the equations enforcing mass and momentum balance about a state in which $\rho_0=0$ and $\bv_0=\bf0$ yields the velocity perturbation equation
\begin{equation}
\label{linmom3bis}
\rho_0 \partial^2_{tt}\bvep =-L\nabla{\rm div}\bvep+\mu \Delta \partial_t \bvep.
\end{equation}
Inserting $\bvep=\bA e^{\sigma t-i\bk\cdot \bx}$ in \eqref{linmom3bis}, we find that
\begin{equation}
\label{linmom4}
(\rho_0\sigma^2 +\mu|\bk|^2\sigma) \bA=L(\bk\cdot\bA)\bA=\tilde M \bA,
\end{equation}
where
\begin{equation}
\tilde M=-L\bk\otimes \bk.
\end{equation}
Since ${\rm det}\tilde M=0$ and ${\rm tr}\tilde M=L|\bk|^2$, the eigenvalues of $\tilde M$ are
\begin{equation}
\tilde\lambda_1= L|\bk|^2
\qquad\text{and}\qquad
\tilde\lambda_2=0.
\end{equation}
The corresponding values of $\sigma$ are
\begin{equation}
\left.
\begin{split}
\displaystyle
\sigma_1(\bk)&=\frac{1}{2\rho_0}
\left(
-\mu|\bk|^2+ \sqrt{ \mu^2 |\bk|^4+4\rho_0 L|\bk|^2}
\right),
\\[4pt]
\sigma_2(\bk)&=\frac{1}{2\rho_0}
\left(
-\mu|\bk|^2- \sqrt{ \mu^2 |\bk|^4+4\rho_0 L|\bk|^2}\right),
\\[4pt]
\sigma_3(\bk)&=-\frac{\mu}{\rho_0}|\bk|^2,
\\[4pt]
\sigma_4(\bk)&=0.
\end{split}\,
\right\}
\end{equation}
For $L>0$, which should be considered generic, the modes corresponding to $\sigma_1$ and $\sigma_2$ are unstable, the mode corresponding to $\sigma_3$ is stable, and the mode corresponding to $\sigma_4$ is neutrally stable.
\section{Conclusions}
\label{sect:conclusions}
%
In this work, the Irving--Kirkwood--Noll procedure is applied to derive the effective meso-scale continuum equations of an active suspension of point particles. The derivations make direct use of the particle equations of motion. A kinetic formulation, often associated with restrictive assumptions of small concentrations and weak interactions, is consequently avoided. The spatially averaged equations enforcing mass and momentum balance are therefore valid for highly concentrated and strongly interacting particle systems. Importantly, we use a realistic model of the self-propulsion force in which the force acting on a particle depends only on the velocity of that particle. This means that in contrast to other agent-based approaches to active suspensions, such as the classical Viscek model, our model does not include a dedicated velocity-aligning mechanism. 

Our approach also involves a novel closure strategy in which the average mass density and velocity are measured not at every point of space-time but rather only at a discrete subset of points. Compared to the standard case of continuum fields defined at each point, upscaling in the discretized setting is associated with an additional loss of information. This makes it necessary to impose certain statistical assumptions about fluctuation tensors of velocities and relative positions. The simplest such assumption is that these tensors are nearly spherical 
and thus can be characterized by scalar parameters reminiscent of the physical temperature. The resulting constitutive theory involves three parameters: the physical temperature and two fluctuation strength parameters. In addition, constitutive equations depend on the mass density and velocity gradient. 
In contrast to previous derivations of continuum models with an ensemble averaging approach \cite{Chuang2007}, the present model includes conservative and dissipative stress tensors, both of which are given by constitutive equations,  and the effects due to fluctuations are taken into account.  

Our coarse-scale evolution equations are similar to the well-known equations of Toner and Tu \cite{toner-tu95}, the main difference being that our equations involve a constitutive relation for the coarse-scale fluctuation force. According to this relation, the strength and the variance of the fluctuation force depends on time and space through the mass density, temperature, and fluctuation strength parameters. 
Previously, kinetic theory has been used to derive Toner--Tu type equations from Vicsek's \cite{vicsek95} model. Since the assumptions underlying classical kinetic theory do not apply to concentrated suspensions, whether the Toner--Tu equations can be reliably applied to such  systems was previously unclear. However, the results of the present work justify the use of these equations for modeling dense active suspensions, at least in the case of nearly spherical particles.

\begin{acknowledgments}
E.F.\ gratefully acknowledges support from the Okinawa Institute of Science and Technology Graduate University with subsidy funding from the Cabinet Office, Government of Japan.
\end{acknowledgments}

\appendix
\section{Closure of conservative stress. Equation of state} 
\label{sect:cons-closure}
\subsection{Taylor approximation of a generic term}

To apply the fluctuation closure to conservative stress (\ref{cons-stress}), we first rewrite $\Psi_\eta$ in \eqref{functPsi} in the form
$$
\Psi_\eta(\bx, \bq_i, \bq_j)=\int_0^1 \psi_\eta \left(\bx-\frac{\bq_i+\bq_j}{2}+(1/2-s)\bq_{ij}\right)\text{d}s.
$$
Recall that $\eta$ is assumed to be much larger than $R$. Then, since $|\bq_{ij}|$ in the stress equation (\ref{cons-stress}) is on the order of the range $R$ of the conservative DPD force $\fbf^C_{ij}$ defined in \eqref{dpd-c-force} and
generic values of $|\bx-(\bq_i+\bq_j)/2|$ are on the order of $\eta\ll R$, we use the Taylor formula centered at $\bx-(\bq_i+\bq_j)/2$ and retain only the first term of the expansion to obtain
\begin{equation}
\label{Psiapp}
\Psi_\eta=\psi_\eta
\left(\bx-\frac{\bq_i+\bq_j}{2}\right) +O\left(\frac{R}{\eta}\right).
\end{equation}
Again, due to the relative smallness of $|\bq_i^\prime|$ and $|\bq_j^\prime|$ in comparison to $\left|\bx-(\overline{\bq}_i+\overline{\bq_j})/2\right|$, we can replace $\bx-(\bq_i+\bq_j)/2$ with $\bx-(\overline{\bq}_i+\overline{\bq_j})/2$ in the leading-order term.
If, as with the particular choice \eqref{psi-box}, the window function $\psi_\eta$ is not differentiable, it can then be approximated by a smooth function and the estimate \eqref{Psiapp} can be applied to that approximation. Alternatively, it is possible to obtain a suitable version of \eqref{Psiapp} directly by noting that the volume-content (or, more technically, the Lebesgue measure) of the set upon which the difference 
$$
\psi_\eta \left(\bx-\frac{\bq_i+\bq_j}{2}+\bigg(\frac{1}{2}-s\bigg)\bq_{ij}\right)-\psi_\eta\left(\bx-\frac{\bq_i+\bq_j}{2}\right)
$$ 
differs from zero is bounded by $cR\eta^{d-1}$, with $c$ being a constant independent of $R$ and $\eta$.

Thus, up to the terms of order $R/\eta$, 
\begin{align}
\bT^C(t, \bx)   &= 
\frac 12 \sum_{i=1}^N \sum_{j=1}^N \fbf^C_{ij} \otimes (\bq_j-\bq_i) \Psi_\eta(\bx, \bq_i, \bq_j)\nonumber\\
& \approx 
-\frac 12 \sum_{i=1}^N \sum_{j=1}^N A w^C (r_{ij}) \frac{\bq_{ij}\otimes \bq_{ij}}{|\bq_{ij}|}\psi_\eta
\left(\bx-\frac{\bar\bq_i+\bar\bq_j}{2}\right)\nonumber\\
& = 
-\frac 12 \frac{1}{\Vbeta}\sum_{(i, j)\in J(t, \bx)} F (r_{ij}) \bq_{ij}\otimes \bq_{ij}, \label{cons-stress-appr} 
\end{align}
where \eqref{dpd-c-force} has been used,  the window function $\psi$ is assumed to be given by \eqref{psi-box}, $r_{ij}=|\bq_{ij}|$,  
\begin{equation}
\label{typical1-1}
F(s)=A\frac{w^C(s)}{s},
\end{equation}
$\bar\bq_i$ and $\bar\bq_j$ are reference positions compatible with the average deformation, 
the index set $J$ for summation is defined by 
\begin{equation}
\label{j-beta}
(i, j)\in J(t, \bx)
\quad \Longleftrightarrow \quad
\frac{\bar\bq_i+\bar\bq_j}{2}\in {\mathcal C}_\bx,
\end{equation}
and ${\mathcal C}_\bx$ denotes the cube with side length $\eta$ centered at $\bx$. 
 
Next, we use Taylor expansions
truncated to the second order in fluctuations, as described in Appendix~\ref{sect:calculations} (see, in particular, \ref{typical4}), to arrive at the result
\begin{multline}
F(r_{ij})\bq_{ij}\otimes\bq_{ij}
=F(|\overline{\bq}|_{ij})(\overline{\bq}_{ij}\otimes\overline{\bq}_{ij}
+\bq^\prime_{ij}\otimes\bq^\prime_{ij})
+ \frac{F^\prime(|\overline{\bq}_{ij}|) }{|\overline{\bq}_{ij}|}
\left[(\bq^\prime_{ij}\otimes\bq^\prime_{ij})(\overline{\bq}_{ij}\otimes\overline{\bq}_{ij})+(\overline{\bq}_{ij}\otimes
\overline{\bq}_{ij})
(\bq^\prime_{ij}\otimes\bq^\prime_{ij})
\right]
\\
\qquad+
\frac 12 \left(F^{\prime\prime}(|\overline{\bq}_{ij}|)\frac{\overline{\bq}_{ij}\otimes \overline{\bq}_{ij}}{|\overline{\bq}_{ij}|^2}+
F^\prime(|\overline{\bq}_{ij}| )\frac{|\overline{\bq}_{ij}|^2\bI-\overline{\bq}_{ij}\otimes\overline{\bq}_{ij}}{|\overline{\bq}_{ij}|^3}
\right): (\bq^\prime_{ij}\otimes \bq^\prime_{ij}) (\overline{\bq}_{ij}\otimes\overline{\bq}_{ij})
+\bR_{ij},
\label{trunc-taylor3}
\end{multline}
where $F^\prime$ and $F^{\prime\prime}$ denote the first and second derivatives of $F$, as defined in \eqref{typical1-1}, with respect to its argument $s$. The last term $\bR_{ij}$ on the right-hand side of \eqref{trunc-taylor3} includes the sum of the first-order terms in $\bq^\prime_{ij}$, which, in view of \eqref{zero-av-fluct}, averages to zero, and higher-order contributions from positional fluctuations.

\subsection{Closed-form conservative stress}

To obtain a closed-form approximation of the conservative stress starting from \eqref{trunc-taylor3},
%
we take advantage of the periodicity of $\overline{\bq}_{ij}$ 
by summing separately all terms with the same value of $\overline\bq_{ij}$.  With this in mind, we let $J_{\alpha\beta}$ denote the index set of $(i, j)$ such that
$\overline{\bq}_{ij}\in {\mathcal C_\beta}$ take the same value $\overline{\bq}^{\alpha\beta}$, and let 
\begin{equation}
\label{cell-strain}
\overline{\bQ}{}^{(\alpha\beta)}=\overline{\bq}^{\alpha\beta}\otimes \overline{\bq}^{\alpha\beta}.
\end{equation}
%
Next, we express the sum in \eqref{cons-stress-appr} as
$$
\sum_{(i,j)\in J_\beta} \cdots =\sum_{\alpha}\sum_{(i, j)\in J_{\alpha\beta}} \cdots.
$$
Each inner sum over $J_{\alpha\beta}$ involves terms typical of those appearing in \eqref{trunc-taylor3}. Factoring out the quantities depending on $\overline{\bq}^{\alpha\beta}$ (which are the same for all terms in that inner sum), we observe that the summation actually applies only to the fluctuation-dependent terms. The result \eqref{trunc-taylor3} further shows that the fluctuations enter the sum over $J_{\alpha\beta}$ through the average fluctuation tensors $\bQ^{\prime, {(\alpha\beta)}}$ defined in \eqref{av-fluct-tensor}.

Granted that $\nbeta$ is sufficiently large, the simplest reasonable assumption that can be made regarding the fluctuations is that they are distributed the same way for all lattice values $\overline{\bq}^{\alpha\beta}$. With this assumption, $\bQ^{\prime,{(\alpha\beta)}}$ is independent of $\alpha$. To further simplify equations, we impose the stronger assumption
\begin{equation}
\label{quad-fluct-assumption}
\bQ^{\prime, {\alpha\beta}}=|\overline{\bq}|^2_\beta \widehat{\bQ}{}^\prime,
\end{equation}
where $\widehat{\bQ}{}^\prime$ is independent of $\beta$. Roughly, this means that, after rescaling by the average volumetric deformation, the distribution of fluctuations is uniform from cell to cell. 

%

Because of the dyadic structure of $\widehat\bQ^\prime$, isotropy is likely to favor the spherical  form
\begin{equation}
\label{appr-spherical}
\widehat\bQ^\prime\approx \theta_q \bI,
\end{equation}
 where $\theta_q$ is a temperature-like parameter that characterizes strength of positional fluctuations. A more detailed explanation of the argument leading to \eqref{appr-spherical} is provided in Appendix~\ref{sect:dyadics}.

Combining \eqref{trunc-taylor3}--\eqref{quad-fluct-assumption}, neglecting the contributions of $\bR_{ij}$, and also using \eqref{appr-spherical}, we arrive at the closed-form approximation,
\begin{multline}
\label{cons-stress-appr1}
\bT^C(\xbeta)\approx \overline{\bT}^C(\xbeta)=
-\frac 12\frac{1}{\Vbeta} 
\sum_\alpha F(|\overline{\bq}^{\alpha\beta}|) \left(\overline{\bQ}{}^{\alpha\beta}+\theta_q \bI\right)
-\frac{\theta_q}{\Vbeta} 
\sum_\alpha  \frac{F^\prime(|\overline{\bq}^{\alpha\beta}|)}{|\overline{\bq}^{\alpha\beta}|} \overline{\bQ}{}^{\alpha\beta}
%
\\-
\frac 14 \frac{\theta_q}{\Vbeta} 
\left[
\sum_\alpha \frac{F^{\prime\prime}(|\overline{\bq}^{\alpha\beta}|)}
{|\overline{\bq}^{\alpha\beta}|^2}{\rm tr}(\overline{\bQ}{}^{\alpha\beta}) \overline{\bQ}{}^{\alpha\beta}+
\sum_\alpha
\frac{F^\prime(|\overline{\bq}^{\alpha\beta}|)}{|\overline{\bq}^{\alpha\beta}|^3}
{\rm tr} (|\overline{\bq}^{\alpha\beta}|^2\bI -\overline{\bQ}{}^{\alpha\beta}){\overline{\bQ}}^{\alpha\beta}
\right],
\end{multline}
of the conservative stress. Since, by \eqref{av-positions}, the relative position vectors $\overline{\bq}^{\alpha\beta}$ depend linearly on the deformation gradient $\nabla\bchi(t, \xbeta)$, \eqref{cons-stress-appr1} expresses $\overline{\bT}{}^C$ as a function of $\nabla \bchi$ and $\theta_q$. To properly interpret \eqref{cons-stress-appr1}, it is important to bear in mind that $F$ is related to the conservative DPD force 
via \eqref{typical1-1}.

\subsection{Equation of state}

For a nearly incompressible fluid, it is to be expected that, up to a rigid rotation, a local deformation gradient is close to a uniform expansion (or contraction). Granted the validity of this expectation, the amounts of stretch in all lattice directions can be assumed to be equal, so that 
all deformed edge vectors have approximately the same length. The expansion (or contraction) of the relative lattice vectors $\bq^{\alpha\beta}$ can be described by a single length scale $|\bq|_\beta$ defined in \eqref{local-length}. Moreover, the conservative stress tensor is approximately equal to $-P\bI$, and the constitutive equation \eqref{cons-stress-appr1} should reduce to an equation of state for the pressure $P$. To deduce the relevant equation of state, observe that summation over $\alpha$ can be interpreted as summation over a cluster of sites that are adjacent to a given site and lie within the range $R$ of the conservative force. This summation can therefore be performed over interaction shells. The first shell contains the closest sites at a distance $|\overline\bq|_\beta l_1$, the second shell consists of sites further away, characterized by the distance $|\overline\bq|_\beta l_2$, and so on. Lattice isotropy implies that all $\bq^{\alpha\beta}$ within the same shell have equal length and, moreover, that the end points of these vectors coincide with the vertices of a regular polyhedron. From symmetry considerations we expect that the sum of dyadic products of these vectors is a spherical tensor,
\begin{equation}
\label{diagon}
\sum_{{\rm shell} \gamma} \overline\bq^{\alpha\beta}\otimes\overline{\bq}^{\alpha\beta}
= C_d |\overline\bq|_\beta|^2 l_\gamma^2 \bI,
\end{equation}
where the constant $C_d$ depends only on the lattice geometry and the dimension $d$ of the physical space. A proof in the two-dimensional case is provided in Appendix~\ref{sect:dyadics}.

Inserting \eqref{diagon} into \eqref{cons-stress-appr1}, we obtain
%
\begin{equation}
\label{cons-split}
\overline{\bT}^C(\xbeta)  \approx  -P(|\overline\bq|_\beta, \theta_q)\bI,
\end{equation}
where the pressure $P$ is given by the equation of state
\begin{multline}
\label{cons-pressure}
P(|\overline\bq|_\beta, \theta_q)=
\frac{1}{{2\mathcal V}_\beta} 
(C_d+\theta_q)|\overline\bq|_\beta^2 \left(\sum_\gamma F(l_\gamma |\overline\bq|_\beta )\right) 
+
\frac{1}{\Vbeta} C_d |\overline\bq|_\beta^3 \theta_q
 \left(\sum_\gamma F^\prime (l_\gamma |\overline\bq|_\beta)l_\gamma^2\right) 
\\
+\frac{3}{{4 \mathcal V}_\beta} C_d^2|\overline\bq|_\beta^4 \theta_q
\left(\sum_\gamma F^{\prime\prime} (l_\gamma |\overline\bq|_\beta)l_\gamma^2\right) 
+\frac{3}{{4 \mathcal V}_\beta} C_d^2|\overline\bq|_\beta^3 \theta_q
\left(\sum_\gamma F^{\prime} (l_\gamma |\overline\bq|_\beta)
\frac{1-C_d l_\gamma^2}{l_\gamma}\right).
\end{multline}
Using 
(\ref{typical1-1}) in (\ref{cons-pressure}), we obtain an expression for the pressure in terms of the conservative DPD forces (\ref{dpd-c-force}), the average deformation $|\bq|_\beta$, and the strength $\theta_q$ of the positional fluctuations.

\subsection{Discussion}
The lattice length $|\overline\bq|_\beta$ can expressed in terms of the average mass density $\overline{\rho}$ using
(\ref{length-density}).  The resulting equation of state gives the pressure as a nonlinear function of $\overline{\rho}$ and the strength $\theta_q$ of positional fluctuations. The role of this parameter is analogous to that of the temperature in the more conventional constitutive models.

A number of additional assumptions were needed to derive the equation of state \eqref{cons-pressure} from the closed-form approximation \eqref{cons-stress-appr1}. It is therefore important to emphasize that \eqref{cons-stress-appr1} is already a bona fide constitutive equation. Indeed, to calculate the stress from this equation, it is only necessary to know the local average deformation $\bchi$. The average fluctuation tensor $\bQ^\prime$ can be then calculated form the initial value of $\hat\bQ{}^\prime$ and $|\overline\bq|_\beta$. Finally, the tensors $\overline{\bQ}{}^{\alpha\beta}$ depend only on the known lattice geometry and $|\overline\bq|_\beta$. Explicit knowledge of the microscopic (DPD) state is consequently \emph{not} needed to calculate the stress. The reductions mentioned in the beginning of this paragraph are motivated by the desire for simplicity and the expectation that the continuum model should be fluid-like. 

It is also important to recognize that distortions in the geometry of the relative vectors $\overline\bq_{ij}$, $i,j=1,2,\dots,N$, that occur near the cell boundary were neglected in going from \eqref{cons-stress-appr1} to \eqref{cons-pressure}.
Indeed if $\overline\bq_i$ is located in $\Cbeta$ and 
$\overline{\bq}_j$ lies in a neighboring cell ${\mathcal C}_{\beta^\prime}$ then the vector $\overline\bq_{ij}$ generally differs from the relative lattice vectors of two points lying within the same cell. This occurs because the lengths and orientations of the lattice vectors of different cells may change after deformation. 
Once again, it is possible to take this effect into account without compromising the structure of the constitutive equation. Although the expressions that would result from doing so would necessarily be more complex than \eqref{cons-stress-appr},  the associated stress would still depend only on the state of deformation. However, this dependence would be non-local in the sense 
that it would incorporate the lattice lengths of not only a local cell but also in all adjacent cells.

It is important to observe that the exact conservative stress has a purely conservative closed-form approximation. This occurs because the exact conservative stress given by \eqref{cons-stress} need not be dissipative. Indeed, 
the fluctuation theorem of Evans, Cohen, and Morriss \cite{ecm93} implies that the second law of thermodynamics may fail 
for molecular dynamical systems driven by conservative forces which are similar to the DPD conservative forces \eqref{dpd-c-force}.
Probability of failure increases with decreasing system size.  This means that some initial conditions may produce dynamical trajectories with decreasing entropy. A related theorem of Gallavotti and Cohen \cite{gc95} guarantees that the time-averaged dynamics are dissipative for almost all initial conditions. The time interval over which averaging is performed must be infinite, and the dynamics, in addition, must be strongly chaotic in a certain suitable sense. 

Consequently, if spatial averaging alone is employed, then the exact conservative stress may not have an accurate dissipative approximation, at least for certain initial conditions. Incorporating time averaging on {\it finite} intervals is still insufficient to ensure that the second law holds. Since a practical theory cannot include impossible to compute long-time averages, we have opted for a method that produces an accurate approximation of the exact stress (provided, of course, that Assumptions 1 and 2 in Section~\ref{empiricalstatistics} hold) without necessarily enforcing dissipativity. The obtained approximation will remain accurate regardless of whether the exact stress is dissipative.

%
\section{Closure of dissipative stress}
\label{sect:diss-closure}
\subsection{Approximation of the generic term}
Consider the exact dissipative stress $\bT^D$ given by \eqref{diss-stress}.  The first step in deriving a closed-form approximation is to invoke the assumption
\begin{equation}
\Psi_\eta(\bx, \bq_i, \bq_j)\approx \psi_\eta\left(\bx,  (\overline{\bq}_i+\overline{\bq}_j)/2\right).
\end{equation}
justified in the beginning of Section \ref{sect:cons-closure}.  
Then, up to the terms of order $R/\eta$, 
\begin{align}
\bT^D(t, \bx)   &= 
\frac 12 \sum_{i=1}^N \sum_{j=1}^N \fbf^D_{ij} \otimes (\bq_j-\bq_i) \Psi_\eta(\bx, \bq_i, \bq_j)\nonumber\\
& \approx 
-\frac 12 \sum_{i=1}^N \sum_{j=1}^N(- \gamma w^D (r_{ij}))(\bv_{ij}\cdot\bq_{ij}) \frac{\bq_{ij}\otimes \bq_{ij}}{r_{ij}^2}\psi_\eta
\left(\bx-\frac{\bar\bq_i+\bar\bq_j}{2}\right)\nonumber\\
& = 
-\frac 12 \frac{1}{\Vbeta}\sum_{(i, j)\in J(t, \bx)} G (r_{ij})(\bv_{ij}\cdot\bq_{ij}) \bq_{ij}\otimes \bq_{ij}, \label{diss-stress-appr} 
\end{align}
where \eqref{diss-force} has been used,  $r_{ij}=|\bq_{ij}|$,  $\psi_\eta$ is given by \eqref{psi-box}, 
\begin{equation}
\label{typical1-12}
G(s)=-\gamma \frac{w^D(s)}{s^2},
\end{equation}
$\bar\bq_i$ and $\bar\bq_j$ are reference positions compatible with the average deformation, and
the index set $J$ over which summation is performed is defined by \eqref{j-beta}.

The generic term
$
G(r_{ij})(\bv_{ij}\cdot\bq_{ij}) \bq_{ij}\otimes \bq_{ij}
$
in \eqref{diss-stress-appr}
can be approximated using Taylor's formula as with the derivation of (\ref{trunc-taylor3}). The result is given by \eqref{typical5}, which contains a large number of terms, some of which can be neglected on the basis of the following consideration. 

Granted that the DPD model is ergodic, spatiotemporal averages can be approximated by ensemble averages. The invariant probability distribution associated with DPD is Gibbs canonical equilibrium distribution 
\begin{equation}
\frac{e^{-\beta H}}{Z},
\label{Gibbsdist}
\end{equation}
where $\beta$ is the inverse temperature, $H=K+U$ is the sum of the kinetic energy $K$ and the potential energy $U$ of the conservative forces, and $Z$ is the normalization constant, traditionally called the partition function. 
This implies that
\begin{equation}
\label{indep}
\overline{\bv^\prime\otimes\bq^\prime}\approx \langle \bv^\prime\otimes\bq^\prime\rangle=
\langle \bv^\prime\rangle \otimes \langle \bq^\prime\rangle
\approx \overline{\bv^\prime}\otimes \overline{\bq^\prime}=\bf0,
\end{equation}
where angular brackets denote ensemble averaging.  Two approximate equalities in \eqref{indep} hold because of assumed ergodicity, the first exact equality holds because the exponential term in \eqref{Gibbsdist} yields a product of the velocity-dependent factor $\exp(-\beta K)$ and a position-dependent factor $\exp(-\beta U)$, and the last exact equality holds because spatial averages of all fluctuations vanish by assumption.

As a consequence of \eqref{indep}, the terms containing products of $\bq^\prime$ and $\bv^\prime$ in the expansion of
\begin{equation}
G(|\bq|_{ij})(\bv_{ij}\cdot\bq_{ij})\bq_{ij}\otimes\bq_{ij}
\end{equation}
provided in \eqref{typical5} can be neglected.  Next, applying straightforward linear algebra to the terms containing $\overline{\bv}_{ij}$ and $\bq_{ij}^\prime$, and temporarily omitting subscripts $(i,j)$ for notational simplicity, we may rewrite \eqref{typical5} as
%
\begin{align}
\label{typical6-c}
G(|\bq|) (\bv\cdot\bq) \bq\otimes \bq&=
G(|\overline{\bq}|) 
\left[
(\overline{\bv}\cdot\overline{\bq})
\left(
\overline{\bq}\otimes \overline\bq+
\bq^\prime\otimes\bq^\prime
\right)
+ 
(\overline\bq\otimes\overline\bv)(\bq^\prime\otimes\bq^\prime) + (\bq^\prime\otimes\bq^\prime) (\overline\bv\otimes\overline\bq) 
\right]
\notag\\
&\quad+\frac 12 G^\prime(|\overline{\bq}|) \frac{1}{|\overline{\bq}|}
\left[
(\overline{\bv}\otimes\overline\bq+\overline{\bq}\otimes\overline{\bv}):(\bq^\prime\otimes\bq^\prime)
\right]
(\overline{\bq}\otimes \overline\bq)
\notag\\
&\quad
+G^\prime(|\overline{\bq}|) \frac{1}{|\overline{\bq}|}(\overline{\bv}\cdot \overline{\bq})
((\bq^\prime\otimes\bq^\prime) (\overline{\bq}\otimes \overline\bq)+ (\overline{\bq}\otimes \overline\bq)(\bq^\prime\otimes\bq^\prime) )
\notag\\
&\quad
+\frac 12  (\overline{\bv}\cdot\overline{\bq}) 
\left[
\left(G^{\prime\prime}(|\overline{\bq}|)\frac{\overline{\bq}\otimes \overline{\bq}}{|\overline{\bq}|^2}+
G^\prime(|\overline{\bq}| )\frac{\bI |\overline{\bq}|^2-\overline{\bq}\otimes\overline{\bq}|}{|\overline{\bq}|^3}
\right): (\bq^\prime\otimes\bq^\prime)  
\right]
(\overline{\bq}\otimes \overline\bq)
+\bR,
\end{align}
where $\bR$ comprises all terms that average to zero and all higher-order terms. In \eqref{typical6-c}, the parenthesis and brackets indicate the order in which various products should be performed. Moreover, $\bA\!:\!\bB=A_{ij} B_{ij}$ denotes the Euclidean inner product of matrices $A$ and $B$. The products of dyadic matrices such as $(\overline\bq\otimes \overline\bq)(\bq^\prime\otimes \bq^\prime)$ are taken in the order in which they are listed, viz.
\begin{align*}
\left[(\overline\bq\otimes \overline\bq)(\bq^\prime\otimes \bq^\prime)\right]_{ij}
& =(\overline\bq\otimes \overline\bq)_{il}(\bq^\prime\otimes \bq^\prime)_{lj}=\overline q_i\overline q_l q^\prime_l q^\prime_j
=(\overline\bq\cdot\bq^\prime) (\overline\bq\otimes\overline\bq^\prime)_{ij}.
\end{align*}


Next, since all particles located in $\Cbeta$ have the same average velocity, $\overline{\bv}_{ij}$ is zero whenever particles $i$ and $j$ lie in $\Cbeta$. Thus, the only nontrivial contributions stem from pairs such that one particle lies in $\Cbeta$ and the other particle lies in an adjacent cell ${\mathcal C}_{\beta^\prime}$. For such pairs, the relative average velocity can be approximated in terms of the average velocity gradient
\begin{equation}
\label{vel-grad}
\overline\bv_{ij}\approx( \nabla\overline\bv )_\beta\overline{\bq}_{ij}.
\end{equation}
Combining \eqref{vel-grad} with \eqref{typical6-c}, we may represent the principal part of a typical term in the dissipative stress as a product of the fourth-order tensor $\bfm^{(\alpha\beta)}$ and the gradient of the average velocity to yield
\begin{equation}
\label{typical7}
G(|\bq_{ij}|) (\bv_{ij}\cdot\bq_{ij}) \bq_{ij}\otimes \bq_{ij}=\bfm^{\alpha\beta} (\nabla\overline\bv)_\beta+\bR.
\end{equation}
The index $\alpha$ refers to the relative lattice vector $\bq^{\alpha\beta}$ such that $\overline{\bq}_{ij}=\bq^{\alpha\beta}$. 
Importantly. with reference to \eqref{typical8-c}, $\bfm^{\alpha\beta}$ depends only on the function $G(s)$ defined in \eqref{typical1-12} which appears in the definition of the DPD dissipative force (compare \eqref{typical1-12} with \eqref{diss-force}), the first and second derivatives of $G(s)$, and the dyadics $\overline{\bq}_{ij}\otimes\overline{\bq}_{ij}$ and $\bq^\prime_{ij}\otimes\bq^\prime_{ij}$.

\subsection{Viscosity approximation}

To obtain a linearly viscous approximation to the dissipative stress, we must sum over all approximations of the form \eqref{typical7} and use the pairwise  `fine-scale viscosity tensors' defined in \eqref{typical8-c}. Although this is not difficult in principle, the equations that result are quite complicated. This is because each relative position vector $\overline{\bq}_{ij}$ now connects a point in the local lattice of one cell with a lattice point in an adjacent cell. 


It is thus evident that $\overline{\bq}_{ij}$ should be referred to a face $\beta\beta^\prime$ between two adjacent cells. The vectors $\overline{\bq}_{ij}$ corresponding to different faces of $\Cbeta$ generally differ. Thus, $|\overline{\bq}_{ij}|$ depends on two values of the mass density, one in $\Cbeta$ (through $|\overline\bq|_\beta$) and that in an adjacent cell ${\mathcal C}_{\beta^\prime}$ through the corresponding quantity $|\overline\bq|_{\beta^\prime}$. So, if all variations of mass density are taken into account, then the resulting viscosity  depends not only on the local value of the mass density, but also on its gradient and, in general, on all of its higher-order gradients. To render the viscosity a local function of mass density, pressure, and temperature, it is reasonable to seek an approximation in which the viscosity at $\xbeta$ depends on $|\overline\bq|_\beta$ but not on the lattice lengths in the adjacent cells. 

To achieve such an approximation, it suffices to assume that $|\overline{\bq}_{ij}|\approx l_{\alpha}|\overline{\bq}|_\beta$, where the non-dimensional parameter $l_\alpha$ refers to a particular relative lattice vector connecting a lattice site with one of the neighboring sites.
With this assumption in place, summation  over $(i,j)$ can be split the into summation over $J_{\alpha\beta}$ followed by summation over $\alpha$, whereby we arrive at an approximation,
\begin{equation}
\label{diss-stress-viscosity}
\bT^D_\beta\approx \overline{\bmu}^{(\beta)}(\nabla\overline{\bv})_\beta,
\end{equation}
of the dissipative stress, where
\begin{equation}
\label{visc-tensor}
\overline{\bmu}^{(\beta)}=\sum_{r=1}^6 \overline{\bmu}^{(\beta, r)},
\end{equation}
with the fourth-order viscosity tensor $\bmu^{(\beta, r)}$ being defined such that
%
\begin{equation}
\left.
\begin{split}
\overline{\mu}^{(\beta, 1)}_{klpm}& =\sum_{\alpha}G(l_\alpha |\overline{\bq}|_\beta) (\overline{\bQ}{}^{\alpha\beta}\otimes\overline{\bQ}{}^{\alpha\beta})_{klpm}, \\
\mu^{(\beta, 2)}_{klpm}& =(\bQ^\prime \otimes \sum_{\alpha}G(l_\alpha|\overline{\bq}|_\beta)\overline{\bQ}{}^{\alpha\beta})_{klpm},
\\
\mu^{(\beta, 3)}_{klpm}& = 
\sum_{\alpha}
 \frac{G^\prime(l_\alpha|\overline{\bq}_\beta|)}{l_\alpha|\overline{\bq}_\beta|}
\left(\overline{Q}^{\alpha\beta}_{kl}
\overline{Q}^{\alpha\beta}_{ps}
\right)
Q^\prime_{sm},
\\
\mu^{(\beta, 4)}_{klpm}& = 
\sum_{\alpha}
\frac{G^\prime(l_\alpha|\overline{\bq}_\beta|)}{l_\alpha|\overline{\bq}_\beta|}
(Q^\prime_{ks}\overline{Q}^{\alpha\beta}_{sl}+\overline{Q}^{\alpha\beta}_{ks}Q^\prime_{sl})\overline{Q}^{\alpha\beta}_{pm},
\\
\mu^{(\beta, 5)}_{klpm}& =\frac 12 
\sum_{\alpha}
 \frac{G^{\prime\prime}(l_\alpha|\overline{\bq}|_\beta)}{l_\alpha|\overline{\bq}_\beta|^2}
(\overline{\bQ}{}^{\alpha\beta}:\bQ^\prime) (\overline{\bQ}{}^{\alpha\beta}\otimes\overline{\bQ}{}^{\alpha\beta})_{klpm},
\\ 
\mu^{(\beta, 6)}_{klpm}& =\sum_{\alpha}
G(l_\alpha |\overline{\bq}_\beta|) 
(\overline{Q}^{\alpha\beta}_{kp}Q^\prime_{lm}+ Q^\prime_{km}\overline{Q}^{\alpha\beta}_{lp}).
\end{split}
\right\}
\label{typical8}
\end{equation}
In each of the foregoing expressions, summation over $\alpha$ can be decomposed into shell summations. However, the shells are no longer spherical. Instead, each shell sweeps out a cone with solid angle less than but approaching $\pi$ in the limit $\nbeta\to\infty$. However, a quick calculation using spherical coordinates shows that
the integral of $\bnu\otimes\bnu$ over a hemisphere is, like its integral over a sphere, a spherical tensor. This makes it possible to once again justify spherical approximations of $\overline\bQ{}^{\alpha\beta}$ and $\bQ^\prime$ in each shell sum.  The $\gamma$-shell sum of $\overline\bQ{}^{\alpha\beta}$ can be approximated by $D_d |\overline\bq|_\beta l^2\gamma\bI$, with $D_d$ depending on the dimension $d$.
The relations (\ref{quad-fluct-assumption}) and (\ref{appr-spherical}) can therefore still be used to approximate $\bQ^\prime$. Also, since $\overline\bQ{}^{\alpha\beta}$ and $\bQ^\prime$ are symmetric tensors, the components of the viscosity tensor must satisfy
\begin{equation}
\label{mu-sym}
\overline{\mu}_{klpm}=\overline{\mu}_{klmp}.
\end{equation}
Using the above observations, it is possible to express \eqref{diss-stress-viscosity} as 
\begin{equation}
\label{diss-stress-viscosity1}
\overline{\bT}^D(t, \bx)=\overline{\bmu}^{(\beta)}(\overline{\rho}(t, \bx), \theta_d)\be(\overline{\bv})(t, \bx),
\end{equation}
where the dependence on $\overline{\rho}$ arises as a consequence of using \eqref{length-density} to express $|\overline\bq|_\beta$ in terms of the average mass density.

\section{Taylor expansions}
\label{sect:calculations}
In this Section we present the second order Taylor expansions of the typical terms in the equations for $\bT^C$ and $\bT^D$.
Dropping the indices $(i, j)$ for notational simplicity, we express a typical contribution to $\bT^C$ as
\begin{equation}
\label{typical1}
F(|\bq|) \bq\otimes\bq,
\end{equation}
where $F$ is defined in terms of $\fbf^C$ in (\ref{typical1-1}).
To obtain the approximation, we write $\bq=\overline{\bq}+\bq^\prime$ and expand $F$ to the second order in $\bq^\prime$, giving
\begin{equation}
\label{F-taylor}
F(|\bq|)= F(|\overline{\bq}|) +F^\prime(|\overline{\bq}|)\frac{\overline{\bq}}{|\overline{\bq}|}\cdot \bq^\prime
+ 
\frac 12 \left(F^{\prime\prime}(|\overline{\bq}|)\frac{\overline{\bq}\otimes \overline{\bq}}{|\overline{\bq}|^2}+
F^\prime(|\overline{\bq}| )\frac{\bI |\overline{\bq}|^2-\overline{\bq}\otimes\overline{\bq}|}{|\overline{\bq}|^3}
\right): \bq^\prime\otimes \bq^\prime+\cdots.
\end{equation}
Mutliplying \eqref{F-taylor} by 
\begin{equation}
\bq\otimes\bq=\overline{\bq}\otimes\overline{\bq}+\overline{\bq}\otimes \bq^\prime+\bq^\prime\otimes
\overline{\bq}+\bq^\prime\otimes\bq^\prime
\end{equation}
and retaining only terms of up to the second order in $\bq^\prime$, we obtain
\begin{align}
\label{typical2}
F(|\bq|) \bq\otimes \bq& = F(|\overline{\bq}|) \overline{\bq}\otimes\overline{\bq}
+F(|\overline{\bq}|) (\overline{\bq}\otimes{\bq}^\prime+{\bq}^\prime\otimes\overline{\bq})
+
F^\prime(|\overline{\bq}|) \overline{\bq}\otimes\overline{\bq}\frac{\overline{\bq}}{|\overline{\bq}|}\cdot\bq^\prime
+ F(|\overline{\bq}|)\bq^\prime\otimes\bq^\prime
\notag\\
& \qquad
+ 
F^\prime(|\overline{\bq}|) \frac{\overline{\bq}}{|\overline{\bq}|}\cdot\bq^\prime
\left(\overline{\bq}\otimes{\bq}^\prime+\bq^\prime\otimes\overline{\bq}\right)
+
\frac 12 \left(F^{\prime\prime}(|\overline{\bq}|)\frac{\overline{\bq}\otimes \overline{\bq}}{|\overline{\bq}|^2}+
F^\prime(|\overline{\bq}| )\frac{\bI |\overline{\bq}|^2-\overline{\bq}\otimes\overline{\bq}|}{|\overline{\bq}|^3}
\right):\left( \bq^\prime\otimes \bq^\prime\right) (\overline{\bq}\otimes\overline{\bq})\notag \\
& \qquad+\cdots.
\end{align}
Rearranging 
\begin{equation}
\label{typical3}
F^\prime(|\overline{\bq}|) \frac{\overline{\bq}}{|\overline{\bq}|}\cdot\bq^\prime
\left(\overline{\bq}\otimes{\bq}^\prime+\bq^\prime\otimes\overline{\bq}\right)=
F^\prime(|\overline{\bq}|) \frac{1}{|\overline{\bq}|}
\left[(\bq^\prime\otimes\bq^\prime)(\overline{\bq}\otimes\overline{\bq})+(\overline{\bq}\otimes\overline{\bq})
(\bq^\prime\otimes\bq^\prime)
\right]
\end{equation}
and dropping the first-order terms in $\bq^\prime$, which average to zero after summing over $(i,j)$,
yields the final expression
\begin{align}
\label{typical4}
F(|\bq|) \bq\otimes \bq& = F(|\overline{\bq}|) \overline{\bq}\otimes\overline{\bq}
+F(|\overline{\bq}|)\bq^\prime\otimes\bq^\prime 
+ 
F^\prime(|\overline{\bq}|) \frac{1}{|\overline{\bq}|}
\left[(\bq^\prime\otimes\bq^\prime)(\overline{\bq}\otimes\overline{\bq})+(\overline{\bq}\otimes\overline{\bq})
(\bq^\prime\otimes\bq^\prime)
\right]
\nonumber\\
&\qquad +
\frac 12 \left(F^{\prime\prime}(|\overline{\bq}|)\frac{\overline{\bq}\otimes \overline{\bq}}{|\overline{\bq}|^2}+
F^\prime(|\overline{\bq}| )\frac{\bI |\overline{\bq}|^2-\overline{\bq}\otimes\overline{\bq}|}{|\overline{\bq}|^3}
\right): (\bq^\prime\otimes \bq^\prime) (\overline{\bq}\otimes\overline{\bq})+\bR,
\end{align}
where $\bR$ denotes the terms that average to zero and higher order terms in $\bq^\prime$.

Terms containing dissipative forces 
$$
\fbf^D=-\gamma \frac{w^D(|\bq|)}{|\bq|^2} \bv\cdot\bq \bq=G(|\bq|)(\bv\cdot\bq) \bq, 
$$ 
where $G|\bq|)$ is defined in \eqref{typical1-12} can be expanded similarly. For brevity, we give only the final result
\begin{align}
\label{typical5}
G(|\bq|) (\bv\cdot\bq) \bq\otimes \bq &= G(|\overline{\bq}|) (\overline{\bv}\cdot\overline{\bq})\overline{\bq}\otimes\overline{\bq}
+G(|\overline{\bq}|)(\overline{\bv}\cdot\overline{\bq})\bq^\prime\otimes\bq^\prime 
+G^\prime(|\overline{\bq}|) \frac{1}{|\overline{\bq}|}(\overline{\bq}\cdot\bq^\prime)(\overline{\bv}\cdot\bq^\prime)
\overline{\bq}\otimes\overline{\bq}
\nonumber\\
&\quad +
G^\prime(|\overline{\bq}|) \frac{1}{|\overline{\bq}|}(\overline{\bv}\cdot\overline{\bq})
\left[(\bq^\prime\otimes\bq^\prime)(\overline{\bq}\otimes\overline{\bq})+(\overline{\bq}\otimes\overline{\bq})
(\bq^\prime\otimes\bq^\prime)
\right]
+
G(|\overline\bq|)(\overline\bv\cdot \bq^\prime) (\bq^\prime\otimes\overline\bq+\overline\bq\otimes \bq^\prime)\nonumber\\
&\qquad\quad + 
\frac 12 \left(G^{\prime\prime}(|\overline{\bq}|)\frac{\overline{\bq}\otimes \overline{\bq}}{|\overline{\bq}|^2}+
G^\prime(|\overline{\bq}| )\frac{\bI |\overline{\bq}|^2-\overline{\bq}\otimes\overline{\bq}|}{|\overline{\bq}|^3}
\right): (\bq^\prime\otimes \bq^\prime)(\overline{\bv}\cdot\overline{\bq}) (\overline{\bq}\otimes\overline{\bq})\nonumber\\
&\qquad\qquad +
G(|\overline\bq|)(\bv^\prime\cdot\bq^\prime) (\overline{\bq}\otimes\overline{\bq})+
G(|\overline\bq|)\left(\frac{\overline{\bq}}{|\overline{\bq}|}\cdot\bq^\prime\right)
(\bv^\prime\cdot\overline\bq) (\overline{\bq}\otimes\overline{\bq})\nonumber\\
&\qquad\qquad\quad+
G(|\overline\bq|)(\bv^\prime\cdot\overline\bq)(\overline{\bq}\otimes{\bq}^\prime+ \bq^\prime\otimes\overline\bq)
+\bR,
\end{align}

Assuming that (\ref{indep}) holds, from (\ref{typical5}) we neglect all terms containing $\bv^\prime$ and $\bq^\prime$. Next, applying straightforward linear algebra to the terms containing $\overline{\bv}$ and $\bq^\prime$, we rewrite (\ref{typical5}) to obtain
(\ref{typical6-c}).

In the simplified notation used in the present section, the approximation
\begin{equation}
\overline\bv_{ij}\approx( \nabla\overline\bv )\overline{\bq}_{ij}
\end{equation}
corresponds to writing
\begin{equation}
\label{vel-grad-c}
\overline{\bv}\approx (\nabla \overline \bv)\overline\bq.
\end{equation}
Combining this with (\ref{typical6-c}) we can represent a typical term in the dissipative stress as a linear operator (the fourth-order tensor) 
$\bfm$ acting on the gradient of the average velocity:
\begin{equation}
\label{typical7-c}
G(|\bq|) (\bv\cdot\bq) \bq\otimes \bq=\bfm \nabla\overline\bv+\bR,
\end{equation}
where 
\begin{equation}
\label{visc-tensor-c}
\bfm=\sum_{r=1}^6 \bfm^{(r)},
\end{equation}
with $\bfm^{(r)}$ given according to
\begin{equation}
\label{typical8-c}
\left.
\begin{split}
\bfm^{(1)}_{klpm}& = G(|\overline{\bq}|) \left[
(\overline{\bq}\otimes \overline\bq)\otimes(\overline{\bq}\otimes \overline\bq)
\right]_{klpm}, 
\\[4pt]
\bfm^{(2)}_{klpm}& =  G(|\overline{\bq}|) \left[(\bq^\prime\otimes\bq^\prime)\otimes (\overline{\bq}\otimes \overline\bq)
\right]_{klpm}\\[4pt]
\bfm^{(3)}_{klpm}& =  G^\prime(|\overline{\bq}|) \frac{1}{|\overline{\bq}|}(\overline{\bq}\otimes \overline\bq)_{kl}
(\overline{\bq}\otimes \overline\bq)_{ps}(\bq^\prime\otimes\bq^\prime)_{sm}, \\[4pt]
\bfm^{(4)}_{klpm}& =  G^\prime(|\overline{\bq}|) \frac{1}{|\overline{\bq}|}
\left[(\bq^\prime\otimes\bq^\prime)_{ks}(\overline{\bq}\otimes \overline\bq)_{sl}+(\overline{\bq}\otimes \overline\bq)_{ks}(\bq^\prime\otimes\bq^\prime)_{sl}\right]
(\overline{\bq}\otimes \overline\bq)_{pm},\\[4pt]
\bfm^{(5)}_{klpm}& =  \frac 12 G^{\prime\prime}(|\overline{\bq}|) \frac{1}{|\overline{\bq}|^2}
((\overline{\bq}\otimes \overline\bq):(\bq^\prime \otimes \bq^\prime)) 
[(\overline{\bq}\otimes \overline\bq)\otimes
(\overline{\bq}\otimes \overline\bq)]_{klpm},
\\[4pt]
\bfm^{(6)}_{klpm}& =G(|\overline{\bq}|) 
((\overline{\bq}\otimes \overline\bq)_{kp}(\bq^\prime\otimes\bq^\prime)_{lm}+ (\bq^\prime\otimes\bq^\prime)_{km}(\overline{\bq}\otimes \overline\bq)_{lp}).
\end{split}
\,
\right\}
\end{equation}

\section{Evolution of the position fluctuation tensor}
\label{sect:positions-fluct}
We define the average position fluctuation tensor 
by
\begin{equation}
\label{pf1}
\bQ^\prime(t, \bx)=\sum_{i=1}^N \sum_{j=1}^N \bq^\prime_{ij}\otimes\bq^\prime_{ij} \psi\left(\bx-\frac{\bq_i+\bq_j}{2}\right),
\end{equation}
where $\bq_i, \bq_j$ on the right-hand side depend on $t$, but this dependence is suppressed.
Taking the time derivative of \eqref{pf1} gives
\begin{multline}
\partial_t \bQ^\prime = \sum_{i=1}^N \sum_{j=1}^N \bv^\prime_{ij}\otimes\bq^\prime_{ij} \psi\left(\bx-\frac{\bq_i+\bq_j}{2}\right)
  + \sum_{i=1}^N \sum_{j=1}^N \bq^\prime_{ij}\otimes\bv^\prime_{ij} \psi\left(\bx-\frac{\bq_i+\bq_j}{2}\right)\\[4pt]
+ \sum_{i=1}^N \sum_{j=1}^N \bq^\prime_{ij}\otimes\bq^\prime_{ij} \left( -\frac{\bv_i+\bv_j}{2}\right)
 \cdot \nabla \psi\left(\bx-\frac{\bq_i+\bq_j}{2}\right).
 \end{multline}
Granted that velocity and position fluctuations are independent, we may take advantage of assumed ergodicity of the DPD equations to show that the products of various functions of the form $f_1(\bq^\prime) f_2 (\bv^\prime)$ average to zero if at least one of the factors $f_1$ and $f_2$ averages to zero. In the present setting, $f_2=\bv^\prime_{ij}$, which averages to zero. Thus, the first two terms on the right-hand side can be neglected, and the third term can be rewritten as
 \begin{align}
 \sum_{i, j=1}^N \bq^\prime_{ij}\otimes\bq^\prime_{ij} \left( -\frac{\bv_i+\bv_j}{2}\right)
 \cdot \nabla \psi\left(\bx-\frac{\bq_i+\bq_j}{2}\right) 
 &=
 \sum_{i, j=1}^N \bq^\prime_{ij}\otimes\bq^\prime_{ij} \left( -\frac{\overline\bv_i+\overline \bv_j}{2}\right)
 \cdot \nabla \psi\left(\bx-\frac{\bq_i+\bq_j}{2}\right)
 \notag\\[4pt]
 &\approx
  - \overline\bv(\bx) \cdot\nabla
 \left(
\sum_{i=1}^N \sum_{j=1}^N \bq^\prime_{ij}\otimes\bq^\prime_{ij} 
 \psi\left(\bx-\frac{\bq_i+\bq_j}{2}\right)
 \right) 
 \notag\\
 & = 
 -\overline{\bv}(\bx)\cdot \nabla 
 \bQ^\prime(\bx).
\end{align}
Thus the actual $\bQ^\prime$ can be approximated by a solution of 
\begin{equation}
\label{pf2}
\partial_t \bQ^\prime+\overline{\bv}\cdot\nabla \bQ^\prime=0.
\end{equation}
This implies that the material derivative of $\bQ^\prime$ is close to zero and, thus, that $\bQ^\prime$ remains close to uniform if it is initially uniform. This justifies the part of Assumption 2 pertaining to $\bQ^\prime$.


\section{Sums of dyadic products of vectors}
\label{sect:dyadics}
\subsection{Dyadic products of unit vectors}
Many quantities of interest in the above closure construction contain sums of dyadic products of vectors of the same length. We now consider such sums in detail. Without loss of generality we assume that all vectors are of unit magnitude.

\subsubsection{Two dimensions}
Let $\{\bim_1,\bim_2\}$ be a positively-oriented Cartesian basis and consider the unit vectors $\bjm_i=\cos i\theta\bim_1+\sin i\theta\bim_2$ where  $\theta={2\pi}/{n}$ and $n$ is an integer. The objective is to evaluate the matrix
$$
M=\sum_{i=1}^n \bjm_i\otimes \bjm_i.
$$
Since 
\begin{equation}
\bjm_i\otimes \bjm_i=
\left(
\begin{array}{cc}
\cos^2 (i\theta) & \cos(i\theta)\sin (i\theta)\\
\cos(i\theta)\sin (i\theta) & \sin^2 (i\theta) \\
\end{array}
\right)
=
\frac 12 \left(
\begin{array}{cc}
1+\cos(2i\theta) & \sin (2i\theta)\\
\sin (2i\theta) & 1-\cos (2i\theta) \\
\end{array}
\right),
\end{equation}
calculation of $M$ reduces to calculating
$
S_1=\sum_{i=1}^n \sin(2i\theta)
$
and 
$
S_2=\sum_{i=1}^n \cos(2i\theta).
$
Using the standard trigonometric identities 
\begin{equation}
\sum_{i=1}^n \sin i\gamma= \frac 12 \cot\frac{\gamma}{2}-\frac{\cos(n+1/2)\gamma}{2\sin\gamma/2}
\qquad\text{and}\qquad
\sum_{i=1}^n \cos i\gamma= -\frac 12 +\frac{\sin(n+1/2)\gamma}{2\sin\gamma/2}
\end{equation}
with $\gamma={4\pi}/{n}$, we find that
\begin{equation}
S_1=\frac 12 \cot\frac{2\pi}{n}-\frac{\cos(4\pi+\frac{2\pi}{n})}{2\sin \frac{2\pi}{n}}=0,
\end{equation}
and that
\begin{equation}
S_2=-\frac 12 +\frac{\sin(4\pi+\frac{2\pi}{n})}{2\sin \frac{2\pi}{n}}=0.
\end{equation}
Therefore,
\begin{equation}
\label{diag-dyad2d}
M=\frac 12 n \bI.
\end{equation}

\subsubsection{Three dimensions}

Next, consider the dyadic product 
$\bnu\otimes\bnu$ with $|\bnu|=1$ in three spatial dimensions. Choose a positively oriented Cartesian basis $\{\bim_1,\bim_2,\bim_3\}$. In spherical coordinates, $\bnu(\theta,\phi)=\sin\theta\cos\phi\bim_1+\sin\theta\sin\phi\bim_2+\cos\theta\bim_3$ for $\phi\in [0, 2\pi]$ and $\theta\in [0, \pi]$. Thus,
\begin{equation}
\label{3d-unit-dyad}
(\bnu\otimes\bnu)(\theta,\phi)=
\left(
\begin{array}{ccc}
\sin^2\theta\cos^2\phi & \sin^2\theta\sin\phi\cos\phi & \sin\theta\cos\theta \cos\phi \\
 \sin^2\theta\sin\phi\cos\phi & \sin^2\theta\sin^2\phi & \sin\theta\cos\theta \sin\phi \\
\sin\theta\cos\theta \cos\phi  & \sin\theta\cos\theta \sin\phi  & \cos^2 \theta
\end{array}
\right),
\end{equation}
from which it follows that the integral of $\bnu\otimes\bnu$ over the unit sphere is given by
\begin{equation}
\label{spherical-dyad-int}
\int_0^\pi\int_0^{2\pi}(\bnu\otimes \bnu)(\theta,\phi) \sin\theta\,\text{d}\phi\mskip2mu\text{d}\theta=\frac{4\pi}{3}\bI.
\end{equation}
If the unit sphere is decomposed into $J$ pieces $U_j$ of equal area 
$$
A=\frac{4\pi}{J}=A(U_j)=\Delta\theta\Delta\phi \sin\theta_j,
$$
then
the sum 
$\sum_{j=1}^J \bnu_j\otimes \bnu_j=\frac{1}{A}\sum_{j=1}^J \bnu_j\otimes \bnu_j A(U_j)$ is a discretization of 
$\frac{1}{A}\int \bnu\otimes\bnu\,\text{d}S$.  Consequently, the tensor
$\sum_{j=1}^J \bnu_j\otimes \bnu_j$ will be approximately spherical. This shows that a sum of a large number of uniformly distributed dyadic products of the form $\bnu_j\otimes \bnu_j$ tends to be spherical. 

From (\ref{3d-unit-dyad}), it also follows that
\begin{equation}
\label{spherical-diagonal1}
(\bnu\otimes\bnu)(\theta, \phi)+(\bnu\otimes\bnu)(\theta, -\phi)+(\bnu\otimes\bnu)(\theta, \pi-\phi)+(\bnu\otimes\bnu)(\theta, \pi+\phi)= 
\left(
\begin{array}{ccc}
4\sin^2\theta\cos^2\phi & 0 & 0 \\
0 & 4\sin^2\theta\sin^2\phi & 0\\
0  & 0 & 4\cos^2 \theta\\
\end{array}
\right)
\end{equation}
and that
\begin{equation}
\label{spherical-diagonal2}
(\bnu\otimes\bnu)(\theta, \phi)+(\bnu\otimes\bnu)(\pi-\theta, \phi)+(\bnu\otimes\bnu)(\theta, -\phi)+(\bnu\otimes\bnu)(\pi-\theta, -\phi)=
\left(
\begin{array}{ccc}
4\sin^2\theta\cos^2\phi & 0 & 0 \\
0 & 4\sin^2\theta\sin^2\phi & 0\\
0  & 0 & 4\cos^2 \theta\\
\end{array}
\right).
\end{equation}
These relations show that diagonalization can occur in sums of small number of terms that are correctly positioned on the unit sphere. Any sum that, together with each $\bnu\otimes\bnu$, contains either triple of rotated dyadic products listed in (\ref{spherical-diagonal1}) or (\ref{spherical-diagonal2}) is therefore a spherical tensor.


\begin{thebibliography}{99}

%
%
\bibitem{baskaran-marchetti2009} A. Baskaran and M. C. Marchetti. Statistical mechanics and hydrodynamics of bacterial suspensions. Proc. Nat. Acad. 108 (2009), 15587--15572.

\bibitem{panchenko-jfm} L. Berlyand and A. Panchenko. Strong and weak blow-up of the viscous dissipation rates for concentrated suspensions. J. Fluid Mech. 578 (2007), 1-34.

\bibitem{bertin06} E. Bertin, M. Droz and G. Gr\'egoire. Boltzmann and hydrodynamic description of self-propelled particles. Phys. Rev. E 74 (2006), 022101.

\bibitem{dense1-boek} E. S. Boek, P. V. Coveney,  N. H. W. Lekkerkerker and P. van der Schoot.  Simulating the rheology of dense colloidal suspension using dissipative particle dynamics. Phys. Rev. E  55, (3) (1997), 3124-3133.

\bibitem{dense2-bolint} D. S. Bolintineanu, G. S. Crest, J.B. Lechman, F. Pierce, S. J. Plimpton and P. R. Schunk. Particle dynamics modeling methods for colloid suspensions. Comp. Part. Mechanics, 1 (2014), 321-356.

\bibitem{bricard} A. Bricard, J.-B. Caussin, N. Desreumaux, O. Dauchot, D. Bartolo,
Emergence of macroscopic directed motion in populations of motile colloids. Nature 503, (2013),
95-98. 

\bibitem{capriz-fried-seguin} G. Capriz, E. Fried, and B. Seguin. Constrained ephemeral continua. Rend. Lincei Mat. Appl. 23 (2012), 157�195.

\bibitem{capriz-ephemeral} G. Capriz. On ephemeral continua. Phys. Mesomech. 11 (2008), 285--298. 


\bibitem{Chuang2007} Y. Chuang, M. R. D'Orsogna, D. Marthaler, A. L. Bertozzi, and L. S. Chayes. State transitions and the continuum limit for a 2D interacting, self-propelled particle system. Physica D 232 (2007), 33--47. 



\bibitem{dunkel2013} J. Dunkel, S. Heidenreich, M. B\"ar and R. E. Goldstein. Minimal continuum theories of structure formation in dense active fluids. New J. Phys. 15 (2013), 045016.

\bibitem{dusenbery} D. B. Dusenbery. Living at micro scale: the unexpected physics of being small. Harvard University Presss, Cambridge, MA, 2009.

\bibitem{ecm93} D. J. Evans, E. G. D. Cohen, and G. P. Morriss. Probability of second law violations in shearing steady flows. Phys. Rev. Lett. 71 (1993), 2401--2404.




\bibitem{espanol-warren} P. Espanol and P. Warren. Statistical mechanics of dissipative particle dynamics. Europhys. Lett. 30 (1995), 191--196. 

\bibitem{espanol95} P. Espanol. Hydrodynamics from dissipative particle dynamics. Phys. Rev. E 52 (1995), 1734--1742.


\bibitem{HPKF} D. Hinz, A. Panchenko, T.-Y. Kim and E. Fried. Motility versus 
fluctuations: Mixtures of self-propelled and passive particles. Soft Matter 10 (2014), 9082--9089.

\bibitem{dpd-first} P. J. Hoogerbrugge and J. M. V. A. Koelman. Simulating microscopic hydrodynamic phenomena with dissipative particle dynamics. Europhys. Lett. 19 (3) (1992), 155--160.

\bibitem{gc95} G. Gallavotti, E. G. D. Cohen. Dynamical ensembles in non-equilibrium statistical mechanics. Phys. Rev. Lett. 74 (1995), 2694--2697.
\bibitem{ibele} M. Ibele, T.E. Mallouk, A.Sen. Schooling behavior of light-powered autonomous micromotors in water. Angew. Chem. Int. Edn. 48, (2009), 3308-3312.

 
\bibitem{ihle11} T. Ihle. Kinetic theory of flocking: Derivation of hydrodynamic equations. Phys. Rev. E 83 (2011), 030901 (R).

\bibitem{Kirkwood} I. Irving and J. G. Kirkwood. The statistical theory of transport processes {IV}. {T}he equations of
hydrodynamics. J. Chem. Phys. 18 (1950), 817--829.


\bibitem{Hardy} R. J. Hardy. Formulas for determining local properties in molecular-dynamics simulations: shock
waves. J. Chem. Phys.  76 (1982), 622--628.

\bibitem{koch-sub} D. L. Koch and G. Subramanian. Collective hydrodynamics of swimming microorganisms: living fluids.
Ann. Rev. Fluid Mech. 43 (2011), 637-659.

\bibitem{kudrolli08} A. Kudrolli, G. Lumay, D. Volfson,  L.S. Tsimring, Swarming and swirling in self-propelled polar granular rods. Phys. Rev. Lett. 100,  (2008), 058001.

\bibitem{kudrolli10} A. Kudrolli, Concentration dependent diffusion of self-propelled rods. Phys. Rev. Lett. 104, (2010), 088001. 

\bibitem{dense3-laurati} M. Laurati, K.J. Mutch, N. Koumakis, J. Zausch, C. P. Amann, A. B. Schofield, G. Petekidis, J. F. Brady, J. Horbach, M. Fuchs and S. U Egelhaaf. Transient dynamics of dense colloidal suspension under shear: shear rate dependence. J. Phys. Condens. Matter 24 (2012), 464104 (13 pages).

\bibitem{marchetti-review} M. C. Marchetti, J. F. Joanny, S. Ramaswamy, T. B. Liverpool, J. Prost, M. Rao and J. Aditi Simha. Hydrodynamics for soft active matter. Rev. Mod. Phys. 85 (2013), 1143--1189.


\bibitem{mb} A. I. Murdoch and D. Bedeaux.
 Continuum equations of balance via weighted averages of microscopic quantities,
{\it Proc. Roy. Soc. Lond. A} 445 (1994), 157--179.

\bibitem{mb96}  A. I. Murdoch and D. Bedeaux.
 A microscopic perspective on the physical foundations of continuum mechanics--Part I: macroscopic states, reproducibility, and macroscopic statistics, at prescribed scales of length and time. Int. J. Eng. Sci. 34 (1996), 1111--1129.

\bibitem{mb97} A. I. Murdoch and D. Bedeaux.
A microscopic perspective on the physical foundations of continuum mechanics II: a projection operator approach
to the separation of reversible and irreversible contributions to macroscopic behaviour. Int. J. Eng. Sci. 35
(1997), 921--949.

\bibitem{murdoch07} A. I. Murdoch.
A critique of atomistic definitions of the stress tensor.
J. Elasticity 88 (2007),  113--140.

\bibitem{murdoch-book} A. I. Murdoch. Physical Foundations of Continuum Mechanics. Cambridge University Press, Cambridge, (2012).

\bibitem{narayan06} V. Narayan, N. Menon, S. Ramaswamy, Nonequilibrium steady states in a
vibrated-rod monolayer: tetratic, nematic and smectic correlations. J. Stat. Mech.: Theory Exp. (2006), P01005.

\bibitem{narayan07} V. Narayan, S. Ramaswamy, N. Menon, Long-lived giant number fluctuations in a swarming
granular nematic. Science 317, (2007), 105-108. 


\bibitem{Noll} W. Noll. Die Herleitung der Grundgleichungen der Thermomechanik der Kontinua aus der
statistischen Mechanik. Indiana U. Math. J. 4 (1955), 627--646.

\bibitem{PBG} A. Panchenko, L. L. Barannyk and R. P. Gilbert, Closure method for spatially averaged dynamics of particle chains.
Nonlinear Anal. 12 (2011), 1681--1697.

\bibitem{PT} A. Panchenko and A. Tartakovsky. Discrete models of fluids: spatial averaging, closure, and model reduction. SIAM J. Appl. Math. 74 (2014), 477--515.

\bibitem{BP} L. L. Barannyk and A. Panchenko. Optimizing performance of deconvolution closure for large ODE systems. IMA J. Appl. Math. 80 (2015), 1099--1123. 

\bibitem{PCKB} A. Panchenko, K. Cooper, A. Kouznetsov and L. L. Barannyk. Kinetic equation for spatially averaged molecular dynamics. Submitted. (Available at: arXiv:1401.2456) 

\bibitem{rabani}A. Rabani, G. Ariel and A. Be�er. Collective Motion of Spherical Bacteria. PLoS ONE (2013) 8(12): e83760. doi:10.1371/journal.pone.0083760.

\bibitem{saintillan-shelley2013} D. Saintillan and M. J. Shelley. Active suspensions and their nonlinear models. C. R. Phys. 14 (2013), 497--517.

\bibitem{schaller}V. Schaller, C. Weber, C. Semmrich, E. Frey, A.R. Bausch,. Polar patterns of driven filaments. Nature 467, (2010), 73-77. 


\bibitem{wensink12} H. H. Wensink, J. Dunkel, S. Heidenreich, K. Drescher, R. E. Goldstein,
H. L�wena, and J. M. Yeomans. Meso-scale turbulence in living fluids. Proc. Natl. Acad. Sci. USA 109 (2012), 14308--14313. 

\bibitem{toner-tu95} J. Toner and Y. Tu. Long-range order in a two-dimensional dynamical XY model: how birds fly together. Phys. Rev. Lett. 75 (1995), 4326--4329.

%
\bibitem{vicsek95} T. Vicsek, A. Czirok, E. Ben-Jacob, I. Cohen, O. Shochet. Novel type of phase transition in a system of self-driven particles. Phys. Rev. Lett. 75 (1995), 1226--1229.

%
%
%
%
%
%
%
%
\end{thebibliography}
\end{document}